\documentclass[aps,prb,reprint,groupedaddress,amsmath,amssymb,superscriptaddress]{revtex4-2}
\usepackage[utf8]{inputenc}
\usepackage[english]{babel}
\usepackage{epsfig}
\usepackage{wrapfig}
\usepackage{bbm}
\usepackage[usenames]{color}
\usepackage{array}
\usepackage{times}
\usepackage{bm}
\usepackage[normalem]{ulem}
\usepackage{graphicx}
\usepackage{physics}
\usepackage{svg}
\usepackage{float}
\usepackage{multirow}
\usepackage[breaklinks=true]{hyperref}
\usepackage{xcolor}
\usepackage{amsmath,amsfonts,amssymb}
\usepackage{graphicx}
\usepackage{hyperref}
\usepackage{color}
\usepackage{stackengine}
\usepackage{subfigure}
\usepackage{verbatim}
\usepackage{comment}
\usepackage{stmaryrd} 

\makeatletter
\def\@email#1#2{%
 \endgroup
 \patchcmd{\titleblock@produce}
  {\frontmatter@RRAPformat}
  {\frontmatter@RRAPformat{\produce@RRAP{*#1\href{mailto:#2}{#2}}}\frontmatter@RRAPformat}
  {}{}
}%
\makeatother

\newcommand{\eps}{\varepsilon}

\newcommand{\RNum}[1]{\uppercase\expandafter{\romannumeral #1\relax}}

\begin{document}
\title{Magnetophotonic crystals with antiferromagnetic order}

\author{Elina A. Kokurina}
\altaffiliation{These authors contributed equally to this work}
\affiliation{School of Physics and Engineering, ITMO University, Saint Petersburg 197101, Russia}

\author{Aleksandra V. Otinova}
\altaffiliation{These authors contributed equally to this work}
\affiliation{Faculty of Physics, Lomonosov Moscow State University, Leninskie gori, Moscow 119991, Russia}

\author{Maxim Mazanov}
\affiliation{School of Physics and Engineering, ITMO University, Saint Petersburg 197101, Russia}

\author{Vladimir I. Belotelov}
\affiliation{Faculty of Physics, Lomonosov Moscow State University, Leninskie gori, Moscow 119991, Russia}
\affiliation{Russian Quantum Center, Moscow 121205, Russia}

\author{Maxim~A.~Gorlach}
\email{m.gorlach@metalab.ifmo.ru}
\affiliation{School of Physics and Engineering, ITMO University, Saint Petersburg 197101, Russia}

\begin{abstract}
We investigate a magnetophotonic crystal formed by the pairs of layers with the opposite out-of-plane magnetization. Despite vanishing total magnetization such antiferromagnetic pattern opens photonic bandgap, gives rise to nontrivial topological phases and enables strong cross-polarized light reflection at frequencies inside the gap which can be harnessed for compact polarization-rotating mirrors. We systematically explore optical properties of this structure highlighting fruitful connections to topological photonics and axion electrodynamics.
\end{abstract}

\maketitle

\section{Introduction}

Engineered photonic structures open rich possibilities to manipulate electromagnetic waves and tailor light-matter interactions at micro- and nanoscale. An especially important class of such structures are periodic photonic crystals~\cite{Joann} where the interplay between the lattice periodicity and the material properties gives rise to photonic bandgaps~\cite{Yablonovitch1987} used for the multitude of applications including mirrors~\cite{Semnani2020}, waveguides~\cite{Johnson2000}, inhibited spontaneous emission~\cite{Yablonovitch1987} and topologically protected light routing~\cite{Ozawa2019}. The simplest yet nontrivial example of such kind is a one-dimensional photonic crystal formed by the pairs of dielectric layers with different permittivities.

If the photonic crystal is composed of magneto-optical materials, this opens a prospect to control light-matter coupling by the static magnetic field which is exploited in magnetophotonic crystals  (MPC)~\cite{Belotelov2005,Inoue2007,lyubchanskii2003magnetic,levy2001flat}. In most of such systems, the net magnetization of the unit cell is the key control parameter, determining fundamental magneto-optical phenomena including Kerr and Faraday rotation. MPCs are known to strongly enhance the Faraday effect near photonic bandgap edges owing to the reduced group velocity of light and increased light–matter interaction~\cite{inoue2006magnetophotonic}. The enhancement becomes particularly pronounced in the structures containing magnetic defect layers, where defect-mode resonances simultaneously enable high transmittance and large Faraday rotation~\cite{inoue2006magnetophotonic}. 

These days MPCs remain an active area of research ~\cite{klos2021magnonics, ignatyeva2024asymmetric, krichevsky2024spatially,gold2024gaga,behjooi2024high,hu2025magnetic,hu2025ultra}.
In particular, the interplay between magnetism and photonic confinement gives rise to asymmetric Faraday rotation~\cite{ignatyeva2024asymmetric}.  Nonreciprocal light propagation in MPC can provide a control of the nonequilibrium emission such as scintillations 
\cite{long2024nonreciprocal}. MPCs can also be used for advanced ultrafast control of spins by the femtosecond  pulses~\cite{krichevsky2024spatially}.
 
Conventionally, MPC is understood as a periodic structure consisting of several dielectric materials of different permittivities, with at least one of the materials magnetized in a fixed direction, which results in the nonzero magnetization of the unit cell. In this Article, we put forward a concept of antiferromagnetic MPC whose net magnetization is compensated, but nevertheless local spatially varying magnetization opens photonic bandgap.
As we prove, the bandgaps in the spectrum of such crystal are  topological. 

Recently, this sort of structures was investigated in the metamaterial limit when the lattice period $a$ is much smaller than the wavelength $\lambda$~\cite{shaposhnikov2023emergent,Seidov2025}. Such metamaterials were shown to exhibit emergent axion response as well as photonic analog of the Witten effect~\cite{shaposhnikov2023emergent}. However, the physics of such antiferromagnetic structures in photonic crystal regime remained uncharted. Here, we systematically investigate the properties of such antiferromagnetic MPC and demonstrate sizable cross-polarized reflection of the incident light at frequencies within the bulk bandgap which can be applied for compact polarization-rotating mirrors.


The rest of the Article is organized as follows. In Sec.~\ref{sec:Bulk} we study the dispersion of the bulk modes and retrieve the associated topological invariants. In Sec.~\ref{sec:Refl} we proceed by calculating reflection and transmission coefficients with a special emphasis on polarization rotation. We finally conclude by Sec.~\ref{sec:Disc} discussing potential applications and prospects.

\begin{figure}[b!]
    \centering
    \includegraphics[width=0.5\linewidth]{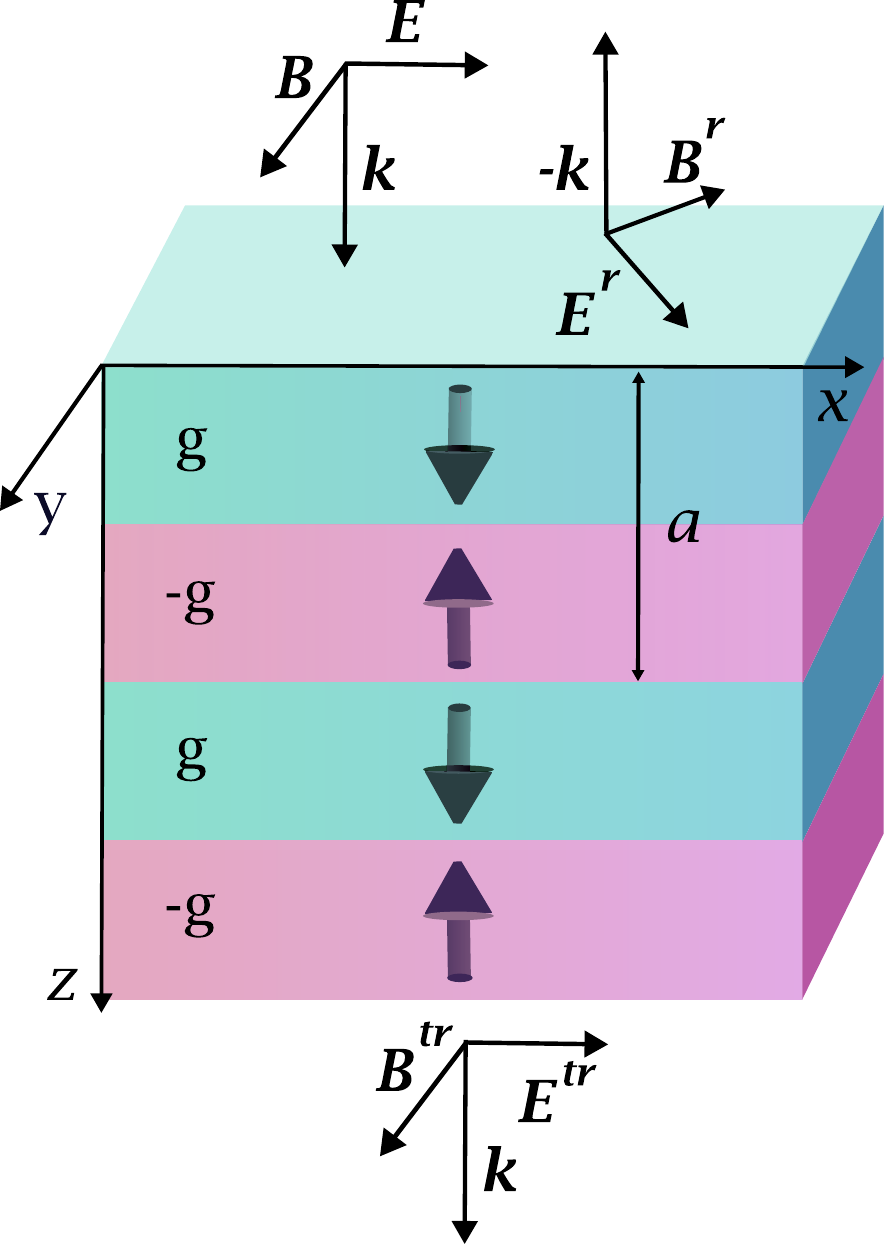}
    \caption{Magnetophotonic crystal consisting of magnetized layers with alternating gyrotropies $g$ and $-g$ and identical permittivity of the layers $\eps$. 
    }
    \label{fig:Slab}
\end{figure}

\section{Bulk dispersion and topology}\label{sec:Bulk}

We start by examining the bulk modes of a one-dimensional (1D) MPC with a period $a$, whose unit cell consists of two ferromagnetic layers of identical thickness $d=a/2$ uniformly magnetized out-of-plane (Fig.~\ref{fig:Slab}). The key feature of this structure is the compensated total magnetization: the gyrotropies of adjacent layers are $g_1=g$ and $g_2=-g$, where $g=fM$, $f$ is the magneto-optical parameter and $M$ is the layer magnetization. The permittivity tensor of each layer reads
\begin{equation}
\hat{\eps}_n=
\begin{pmatrix}
\eps_\alpha && ig_\alpha &&0\\\ -ig_\alpha &&\eps_\alpha&&0\\0&&0&&\eps_\alpha
\end{pmatrix}\:,
\end{equation}
where $\alpha=1,2$ and $\eps_\alpha$ is the diagonal component of the permittivity tensor of the respective layer. For simplicity, the permeability of the layers is set to unity which is the case at optical frequencies.

We utilize the basis of circularly polarized modes described by the vectors
$\mathbf{e}_{\pm}=\mathbf{e}_x\pm i\mathbf{e}_y$, where $+$ and $-$ correspond to right (RCP) and left (LCP) circular polarizations, respectively. In each layer the refractive indices for a given circular polarization $\eta=\pm 1$ for RCP/LCP can be expressed as (see Supplementary Materials, Sec.~1):
\begin{equation}
n_{\alpha}^{\eta}=\sqrt{\eps_{\alpha}-\eta g_{\alpha}}.
\label{ref}
\end{equation}
To analyze the  propagation of a wave with a Bloch wave number $k$ in the periodic structure, we employ the transfer matrix method (Supplementary Materials, Sec.~2). 
The dispersion equations for right and left circular polarizations decouple and take the form:
\begin{align}
\cos(ka)=\cos(k^{\eta}_1d)\cos(k^{\eta}_2d)-\zeta^{\eta}\sin(k^{\eta}_1d)\sin(k^{\eta}_2d)\:.
\label{eq:dispersion}
\end{align}
Here 1 and 2 indices label the layer within the unit cell,  $k^{\eta}_\alpha=q\, n^{\eta}_{\alpha}$, $q=\omega/c$ and $\zeta^{\eta}=\frac{1}{2}\left(\frac{n^{\eta}_2}{n^{\eta}_1}+\frac{n^{\eta}_1}{n^{\eta}_2}\right)$.

If the permittivities of the layers are identical ($\eps_1=\eps_2=\eps$), the finite structure consisting of an integer number of unit cells exhibits $\mathcal{PT}$ symmetry, where $\mathcal{P}$ is spatial inversion and $\mathcal{T}$ is time reversal. The same symmetry applies to the periodic structure. It connects right and left circularly polarized modes and guarantees their degeneracy. This can  also be seen directly from Eq.~\eqref{eq:dispersion} where equal permittivities of the layers result in $\zeta^{(+)}=\zeta^{(-)}=\zeta$.

\begin{figure}
    \centering
    \includegraphics[width=0.9\linewidth]{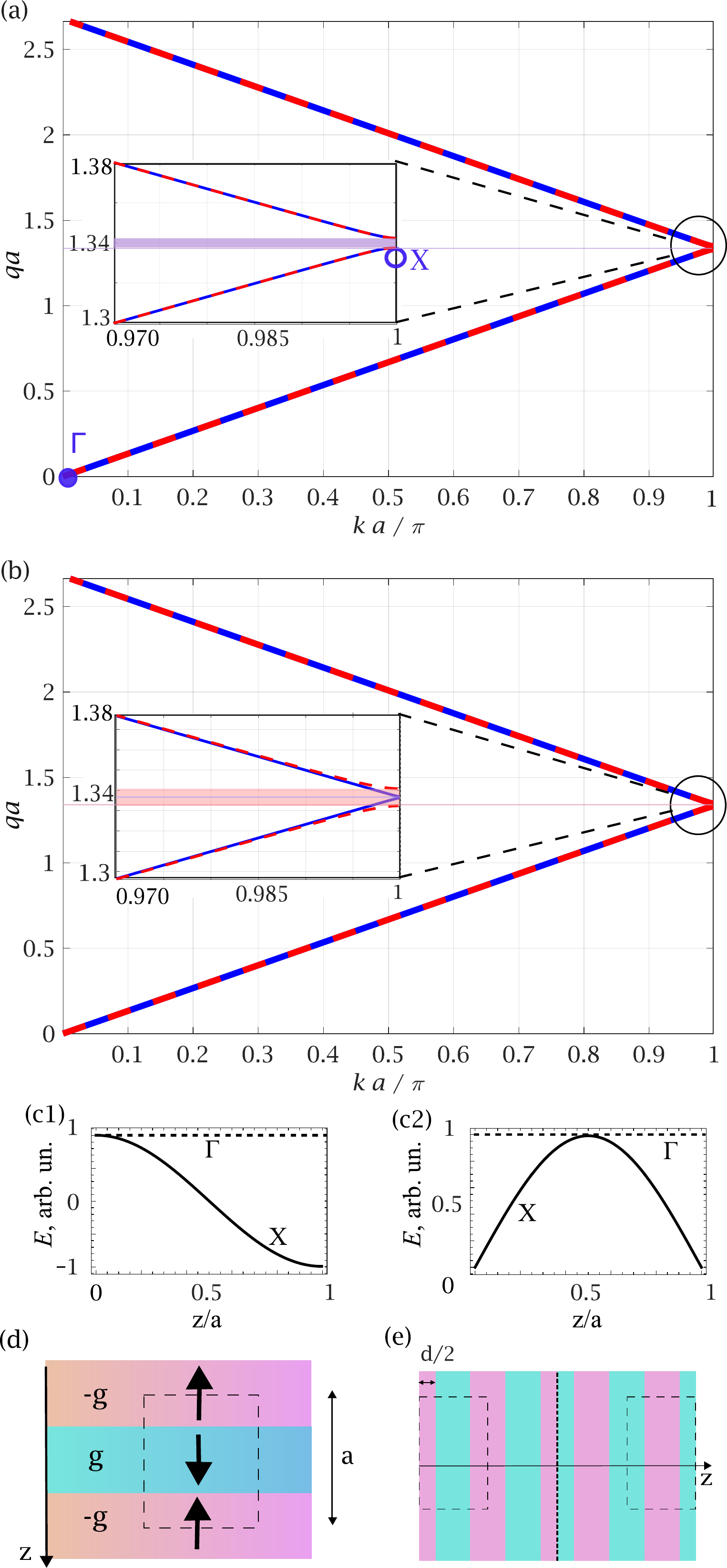}
    \caption{Photonic band structure of antiferromagnetic magnetophotonic crystal. (a) Calculated dispersion for $\varepsilon_1 = \varepsilon_2 = 5.5$, $g = 0.03$ corresponding to bismuth iron garnet films~\cite{kahl2003optical}. The dispersion of RCP and LCP modes coincides. Photonic bandgap is shaded. (b) The band structure for the different layer permittivities $\varepsilon_1 = 5.5$, $\varepsilon_2 = 5.55$. The degeneracy of RCP and LCP modes shown by red and blue lines, respectively, is now lifted. 
    (c) Electric field amplitude distributions in the inversion-symmetric unit cell for the lowest-frequency Bloch modes with RCP~(c1) and LCP~(c2) at points $\Gamma$ ($k=0$, zero frequency, dashed line) and $\rm X$ ($k=\pi/a$, lowest branch, solid line).  (d) Choice of the unit cell for the Zak phase calculation. (e) Interface of two MPCs with the different terminations enabling localized topological mode. }
    \label{fig:Fig2}
\end{figure}

Notably, despite the identical permittivities $\eps_1$ and $\eps_2$, spatially varying magnetization creates spatial periodicity and enables photonic bandgap. If the gyrotropy of the layers is weak enough ($|g|\ll\eps$), which is the case for experimentally available structures, the bandgap is narrow and the center-of-bandgap frequency is calculated as
\begin{equation}
\omega_{B}=\pi c\,/d(n_++n_-)\approx\pi c /(2\sqrt{\eps} d)\:,   
\end{equation}
where $n_{\pm}=\sqrt {\eps\mp g}$ and $c$ is the speed of light in vacuum (see derivation in Supplementary Materials, Sec.~2). Thus, the position of the gap can be tuned by choosing the thickness of the layers. In turn, the width of the photonic bandgap reads
\begin{equation}\label{eq:gap}
\Delta\omega\approx\frac{c|g|}{\eps^{3/2} d}
\end{equation}
%
and is roughly proportional to the gyrotropy $g$ which defines the difference between $n_{+}$ and $n_{-}$ refractive indices. These results are illustrated by the calculated band structure in Fig.~\ref{fig:Fig2}(a).

If permittivity modulation and gyrotropy are present simultaneously, the $\mathcal{PT}$ symmetry of the structure breaks down and the degeneracy of the two circularly polarized modes is lifted. The two mechanisms of the gap opening compete resulting in the different gaps for the different polarizations. In particular, $g=(\eps_1-\eps_2)/2$ closes the bandgap for the RCP polarization. On the other hand, if $g=(\eps_2-\eps_1)/2$, LCP gap closes. The representative band structure for the case of unequal $\eps_1$ and $\eps_2$ is depicted in Fig.~\ref{fig:Fig2}(b).

Besides the dispersion of the bulk modes an important role is played by the topology of the Bloch modes which via bulk-boundary correspondence determines the existence of the localized modes at the interface of the structures with the different topological invariants~\cite{Xiao2014Apr,Ozawa2019}. In the one-dimensional case, the relevant topological invariant is the Zak phase $\theta_n$ of the individual Bloch band~\cite{Xiao2014Apr} which takes quantized values $0$ and $\pi$ for the inversion-symmetric unit cell~\cite{Hughes2011}. 

We retrieve the topology of the lowest band by checking the inversion parities of right- and left-circularly polarized Bloch modes at two high-symmetry points of the Brillouin zone: $\Gamma$ and $\rm X$ with $k=0$ and $k=\pi/a$, respectively. Clearly, electric field profile at the $\Gamma$ point is always symmetric corresponding to the homogeneous field. For RCP mode, the field amplitude distribution at the $\rm X$ point is inversion-odd [Fig.~\ref{fig:Fig2}(c1)] and hence $\theta_1^{\text{RCP}}=\pi$. In contrast, LCP mode features inversion-even field profile at $\rm X$ point [Fig.~\ref{fig:Fig2}(c2)] and therefore $\theta_1^{\text{LCP}}=0$. The respective unit cell choice is shown in Fig.~\ref{fig:Fig2}(d). Note that for a different unit cell choice centered at the layer with gyrotropy $-g$ the Zak phases for RCP and LCP modes interchange.

A direct consequence of the nontrivial Zak phase is the  topological localized mode at the interface between the two MPCs with the opposite values of $g$ illustrated in Fig.~\ref{fig:Fig2}(e). As analyzed in the Supplementary Materials Sec.~4, this in-gap localized mode results in the extra narrow transmission peak and associated reflection minimum within the band gap. Slightly unequal $\eps_1 \neq \eps_2$ yield qualitatively the same result as long as both polarization bandgaps remain open.


\section{Reflection and transmission properties}\label{sec:Refl}

Opening of gyrotropy-induced photonic bandgap has immediate consequences for the optical properties of the structure. First, we analyze a semi-infinite photonic crystal and normal incidence. 
At frequencies within the bulk bandgap all incident light is reflected, i.e. the reflection coefficients are $|r_{+}|=|r_{-}|=1$, where $r_{\pm} = E_{\pm}/E_{\pm}^{(\text{inc})}$ is the amplitude reflection coefficient. 
If the incident wave is linearly polarized, the polarization of reflected wave stays linear, but with the rotated polarization plane. What is less intuitive, the cross-polarized reflection dominates the co-polarized one in the bandgap. In particular, at center-of-bandgap frequency $\omega_B$ the reflection coefficients to the leading order in $g$ read:
\begin{gather}
 r_{xx}(\omega_B) = -i g 
\pi (\varepsilon + 1) / (8 \varepsilon^{3/2})\:,\label{eq:rxxbg}\\
r_{xy}(\omega_B)
=-i - g \pi (\varepsilon - 1) / (8 \varepsilon^{3/2})\:.\label{eq:rxybg}
\end{gather}
%
Thus, for small gyrotropy $g$, the reflected wave is almost fully cross-polarized with small $O(g)$ leakage into the co-polarized reflection,
while polarization rotation is close to $90^\circ$:
\begin{equation}
\theta = \pi/2 + g \pi (\varepsilon + 1))/(8 \varepsilon^{3/2})\:,    
\end{equation}
see Supplementary Materials, Sec.~3 and Supplementary Fig.~S2. Note that these expressions are derived assuming photonic bandgap which ensures full reflection of the incident light and hence they are only valid for small but nonzero $g$.

In addition, if the permittivities of the layers are equal $\eps_1=\eps_2$, the reflected light keeps linear polarization also at frequencies outside the photonic gap, which is due to the fact that the optical impedances $Z_{\pm}=E_{\pm}/H_{\pm}$ are complex conjugated. In contrast, for $\eps_1\ne\eps_2$ the reflected wave acquires ellipticity at frequencies outside the photonic gap.

Another instructive limit is the metamaterial regime~$qa\ll1$ when the wavelength covers multiple lattice periods. In this limit, the expressions for the reflection coefficients can be simplified yielding
\begin{align}
    &r_{xx}=\frac{1-\eps}{\Delta}, \qquad 
    r_{xy}=- \frac{2 \chi}{\Delta}\:, \label{rxyax}
\end{align}
where $\Delta=(1+\sqrt{\eps})^2, \chi=\frac{\pi}{2}g \frac{a}{\lambda}.$
This limit is understood in terms of axion electrodynamics~\cite{shaposhnikov2023emergent}, where $\chi$ quantifies emergent axion response of the metamaterial.

\begin{figure}[b]
    \centering
    \includegraphics[width=0.85\linewidth]{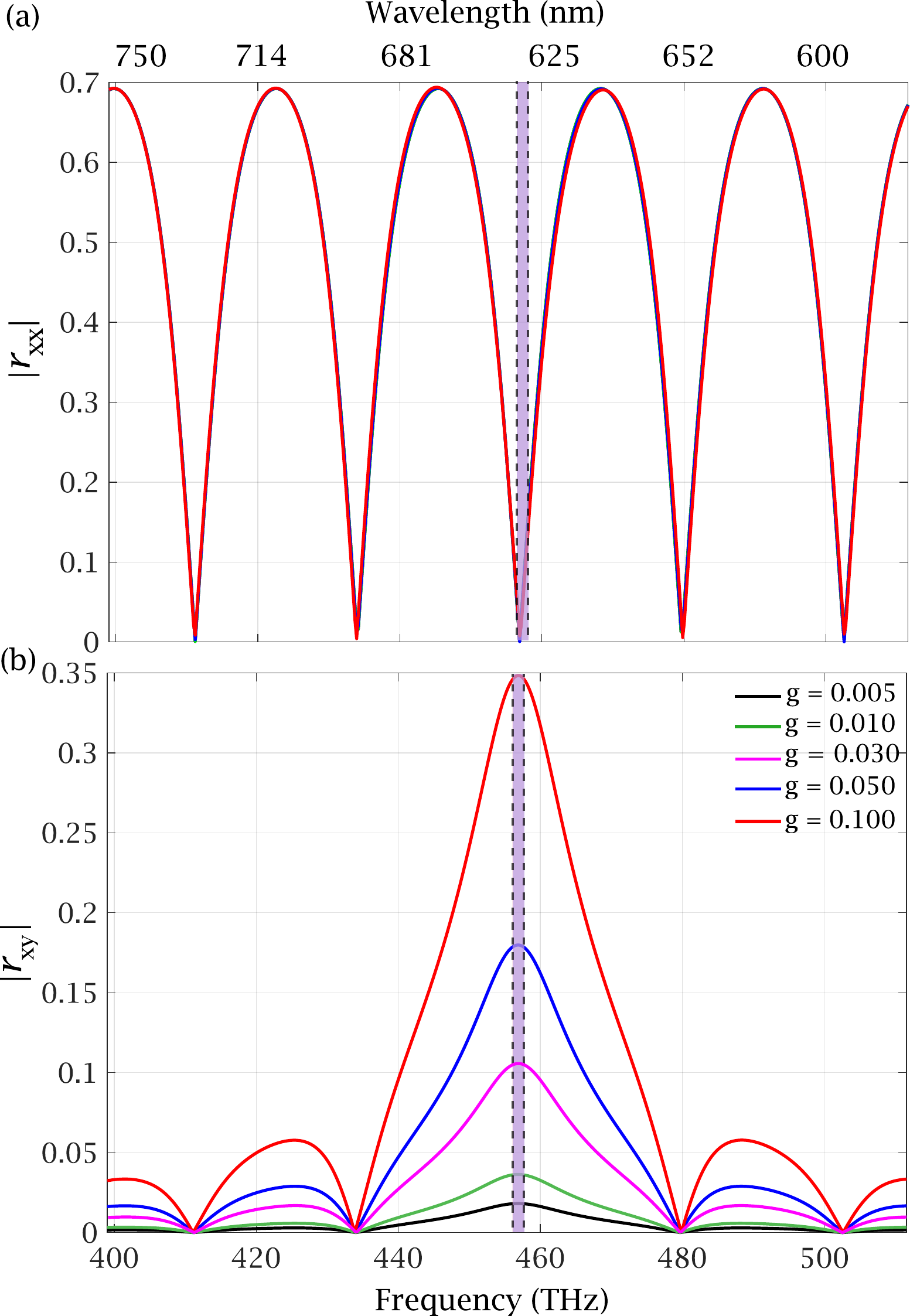}
    \caption{Reflection from a finite slab of antiferromagnetic MPC for the different gyrotropy strengths ranging 0.005 to 0.1. Co-polarized (a) and cross-polarized (b) reflection coefficients are calculated for layer permittivities $\eps_1=\eps_2=5.5$, number of periods $N=20$ and layer thickness $d=70$~nm. Shaded area indicates the bandgap for the maximal gyrotropy.
   }
    \label{fig:gdependency}
\end{figure}

\begin{figure}
    \includegraphics[width=1.0\linewidth]{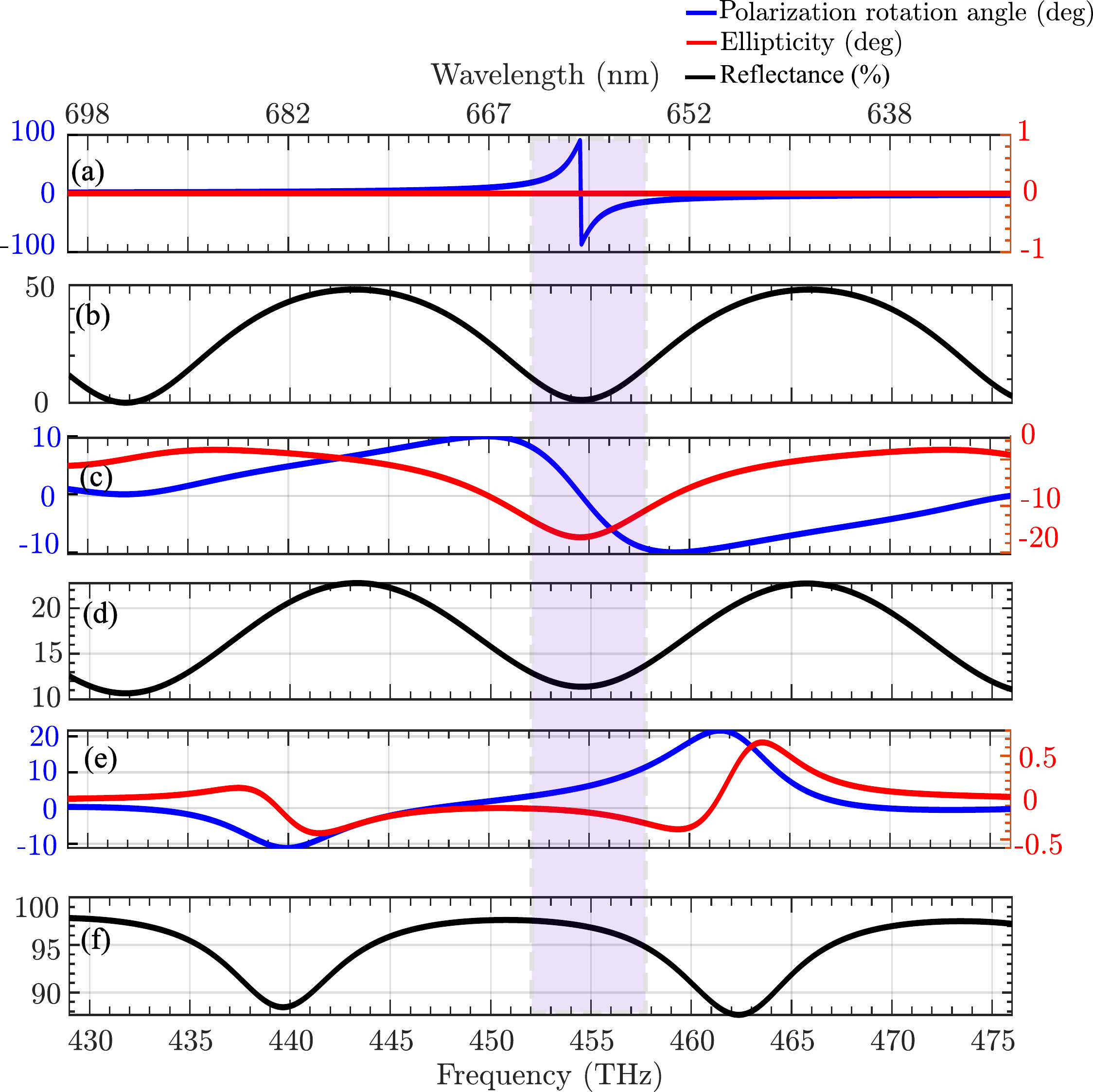}
    \caption{Simulated polarization rotation angle and ellipticity in reflection. Panels (a,b), (c,d), and (e,f) correspond to the different substrates: (a,b) air‑structure‑air (no substrate), (c,d) gadolinium gallium garnet substrate with $\varepsilon$ = 4, and (e,f) silver substrate. All layers of the structure have the same permittivity $\varepsilon$ = 5.5, each layer exhibits $g=\pm 0.03$, number of periods is 20. Ellipticity is zero for the symmetric structure (a), reaches about $-20^{\circ}$ for the substrate with $\varepsilon$ = 4 and drops nearly to $0^{\circ}$ for the silver‑backed structure (e).}       
    \label{fig:far_and_ellipt_same_eps}
\end{figure}

Less trivial is the case of a finite slab consisting of $N$ periods (i.e. $2\,N$ layers) and having the thickness $L=Na$. The typical spectrum of co-polarized reflection in Fig.~\ref{fig:gdependency}(a) features  characteristic oscillatory behavior due to Fabry-Perot resonances of a finite slab. The co-polarized reflection is practically insensitive to the gyrotropy strength [Fig.~\ref{fig:gdependency}(a)].

The behavior of cross-polarized reflection is more remarkable [Fig.~\ref{fig:gdependency}(b)]. While featuring the similar Fabry-Perot resonances, the cross-polarized reflection strongly depends on gyrotropy reaching a sharp maximum at the frequency of the bulk bandgap in agreement with the semi-infinite case. Given that the co-polarized reflection at the same frequency is suppressed, this indicates up to $90^\circ$ polarization rotation of reflected light even for finite-thickness samples. 
{Analytical calculations detailed in Supplementary Materials, Sec.~6 yield the following expressions for the reflection coefficients at center-of-bandgap ($\omega_B$) frequency to the leading order in small parameter $g/\varepsilon \ll 1$:
\begin{eqnarray}
r_{xy}
\simeq 
- i \tanh \frac{N g}{\varepsilon} 
,  \qquad 
r_{xx} \simeq \alpha \frac{g}{\varepsilon} r_{xy}
, 
\end{eqnarray}
where $\alpha = (\pi/8)
\left(
n / n_a + n_a / n
\right)
\sim 1
$, 
$n=\sqrt{\varepsilon}$, $n_a = \sqrt{\varepsilon_a}$ is refractive index of the ambient medium, 
and the corresponding rotation angle reads 
$
\theta
\simeq
\pi/2 
+
\alpha g / \varepsilon
$. 
These results confirm that $|r_{xy}| \gg |r_{xx}|$ at the midgap frequency and are in excellent correspondence with numerical simulations.} 
At the same time, cross-polarized transmission vanishes identically [Supplementary Materials, Sec.~3 and Figs.~S3-S4].

To characterize the polarization of reflected light, we compute the polarization rotation $\theta$ and ellipticity $\eta$ which quantify the alignment of the principal axis of polarization ellipse and the ratio of its principal axes, respectively, and are defined in the Supplementary Materials, Sec.~5.  

First we examine the idealized scenario when a finite MPC is placed in air. Polarization rotation shown in Fig.~\ref{fig:far_and_ellipt_same_eps}(a) indeed reaches extreme values of $90^\circ$ in agreement with  Fig.~\ref{fig:gdependency}.  At the same time, ellipticity vanishes indicating that the polarization of reflected light stays linear. However, the magnitude of reflection is quite low [Fig.~\ref{fig:far_and_ellipt_same_eps}(b)]. This should be attributed to the relatively weak gyrotropy and small number of periods of the finite crystal, which appear insufficient to enhance reflection.

In experiments, the structure is necessarily placed on a substrate. The presence of substrate breaks the $\mathcal{PT}$ symmetry and changes the reflection properties qualitatively. Fig.~\ref{fig:far_and_ellipt_same_eps}(c) shows the case of the same finite MPC placed on a gadolinium gallium garnet (GGG) substrate with $\varepsilon = 4$, commonly used for the magneto-optical film growth. The resonant polarization rotation is now suppressed down to $\theta \approx 10^\circ$ at a higher level of reflectance around 15$\%$, while ellipticity gets rather pronounced around $\eta \approx -10^\circ$ at that frequency. 

In such setting, it is instructive to compare the performance of antiferromagnetic MPC to a single magnetized film of the same thickness deposited on the same GGG substrate. Our calculations (Supplementary Materials, Sec.~7) show that the two systems feature comparable reflectance, while polarization rotation for antiferromagnetic MPC is nearly twice higher than that for the single magnetized film of the same thickness on a dielectric substrate.

The problem of relatively low reflectance can be remedied by replacing the dielectric substrate by a silver mirror, Fig.~\ref{fig:far_and_ellipt_same_eps}(e,f). This boosts the reflectance up to nearly 90\% and increases polarization rotation nearly twice, up to $20^\circ$. Such two-fold enhancement compared to the previous scenario can be connected to the double passage of light through antiferromagnetic structure. The reflection is still below 100\% because of the losses in  silver~\cite{johnson1972optical}. The losses also ensure nonzero ellipticity $\eta\approx -0.5^\circ$, which is substantially suppressed compared to GGG substrate. However, the comparison between antiferromagnetic MPC and a single magnetized film in such mirror-backed configuration appears less favorable, and antiferromagnetic structure does not have advantages in polarization rotation (Supplementary Materials, Sec.~7).


Finally, we estimate the effect of losses, taking the available experimental data for ferrite garnet thin films. The calculations show that moderate losses both in permittivity and gyrotropy leave polarization rotation practically unchanged, while the reflectance drops by few percent (Supplementary Materials, Sec.~8). This confirms that the above results are applicable to the experimental structures if losses are kept at moderate experimentally available level.


\section{Discussion and conclusions}\label{sec:Disc}

In summary, we have put forward the design of antiferromagnetic magnetophotonic crystal where the photonic band gap arises due to the spatially alternating magnetization. Even though the net magnetization is perfectly compensated, the structure exhibits sizable cross-polarized light reflection at bandgap frequencies serving as a compact magnetically controllable polarization-rotating mirror. Such MPC can be implemented experimentally using bismuth rare earth iron garnets of different compositions via pulsed laser deposition and magnetron sputtering~\cite{kahl2003optical}. Realistically, such procedure may yield up to 20-30 lattice periods (40-60 layers) of good quality.


We also demonstrate that the resulting photonic gap is topological giving rise to the localized topological modes at the interface of MPCs with the different gyrotropy signs. However, the localization length of the topological mode is too large to be observed in finite samples with 20 lattice periods.

On the other hand, the low-frequency limit of the same structure realizes the instance of axion electrodynamics~\cite{Nenno2020}, where nonzero cross-polarized light reflection measures the strength of emergent axion field.

These findings feature interesting parallels with X-ray optics where magnetic Bragg cross-polarized reflection peaks serve as a valuable tool~\cite{Belyakov1989Aug} for detecting and imaging of antiferromagnetic domains~\cite{Kim2018Nov} and for separating different individual contributions to magnetic response in  materials~\cite{Gibbs1988Sep,Simeth2023Jun} such as distinct atomic species or valence environments~\cite{Hepting2018Nov}.

Therefore, this work unites together such diverse concepts as magnetophotonic crystals, emergent axion fields and topology and reveals new compact tools for polarization conversion.


\begin{acknowledgments}

Theoretical and numerical models for the periodic antiferromagnetic magnetophotonic crystal were  supported by the Russian Science Foundation, grant No.~26-72-10054. Theoretical analysis of topological properties was supported by Priority 2030 Federal Academic Leadership Program. Numerical simulations of finite antiferromagnetic slabs were supported by the Russian Science Foundation, grant No.~24-42-02008. A.V.O. and V.I.B acknowledge support of the analysis and comparison of the antiferromagnetic and ferromagnetic photonic crystals from the Foundation for the Advancement of Theoretical Physics and Mathematics ``Basis'' grant No.~25-1-1-49-1.

\end{acknowledgments}

\section*{Data Availability}

The data that support the findings of this study are provided within the article and supplementary materials.


\bibliography{PBG}

@Inbook{Inoue2007,
author="Inoue, M.
and Granovsky, A.
and Aktsipetrov, O.
and Uchida, H.
and Nishimura, K.",
editor="Akta{\c{s}}, Bekir
and Mikailov, Faik
and Tagirov, Lenar",
title="Magnetophotonic Crystals",
booktitle="Magnetic Nanostructures",
year="2007",
publisher="Springer Berlin Heidelberg",
address="Berlin, Heidelberg",
pages="29-43",
isbn="978-3-540-49336-5",
doi="10.1007/978-3-540-49336-5_4",
url="https://doi.org/10.1007/978-3-540-49336-5_4"
}

@article{long2024nonreciprocal,
	author = {Long, Olivia Y. and Pajovic, Simo and Roques-Carmes, Charles and Tsurimaki, Yoichiro and Rivera, Nicholas and Solja{\ifmmode\check{c}\else\v{c}\fi}i{\ifmmode\acute{c}\else\'{c}\fi}, Marin and Boriskina, Svetlana V. and Fan, Shanhui},
	title = {{Nonreciprocal scintillation using one-dimensional magneto-optical photonic crystals}},
	journal = {Phys. Rev. Appl.},
	volume = {22},
	number = {5},
	pages = {054062},
	year = {2024},
	month = nov,
	publisher = {American Physical Society},
	doi = {10.1103/PhysRevApplied.22.054062}
}

@article{hu2025ultra,
	author = {Hu, Shiqi and Shen, Chao and Chen, Kaifeng and Chen, Ying and Xiao, Wei and Liu, Gui-Shi and Chen, Lei and Zhou, Guofu and Chen, Zhe and Cao, Donglin and Chen, Yaofei and Luo, Yunhan},
	title = {{Ultra-high FOM magneto-optical surface plasmon resonance sensor based on magnetophotonic crystals and hyperbolic metamaterials}},
	journal = {Measurement},
	volume = {252},
	pages = {117396},
	year = {2025},
	month = aug,
	issn = {0263-2241},
	publisher = {Elsevier},
	doi = {10.1016/j.measurement.2025.117396}
}

@article{hu2025magnetic,
	author = {Hu, Zhijia and Qu, Guangyin and Guo, Lulu and Zhang, Xiaojuan and Li, Siqi and Kuai, Yan and Fu, Weiwei and Gao, Jiangang and Xu, Feng and Xia, Jiangying and Yu, Benli},
	title = {{Magnetic Photonic Crystal for Adaptive Flexible Lasers and Secure Encoding}},
	journal = {Laser Photonics Rev.},
	volume = {19},
	number = {24},
	pages = {e02240},
	year = {2025},
	month = dec,
	issn = {1863-8880},
	publisher = {John Wiley {\&} Sons, Ltd},
	doi = {10.1002/lpor.202502240}
}

@article{behjooi2024high,
	author = {Behjooi, T. and Ghanaatshoar, M.},
	title = {{High performance electrically-derived single-pixel magnetophotonic spatial light modulator}},
	journal = {J. Magn. Magn. Mater.},
	volume = {599},
	pages = {172064},
	year = {2024},
	month = jun,
	issn = {0304-8853},
	publisher = {North-Holland},
	doi = {10.1016/j.jmmm.2024.172064}
}

@article{gold2024gaga,
	author = {Gold, Hannah and Pajovic, Simo and Mukherjee, Abhishek and Boriskina, Svetlana V.},
	title = {{GAGA for nonreciprocal emitters: genetic algorithm gradient ascent optimization of compact magnetophotonic crystals}},
	journal = {Nanophotonics},
	volume = {13},
	number = {5},
	pages = {773--792},
	year = {2024},
	month = mar,
	issn = {2192-8614},
	publisher = {John Wiley {\&} Sons, Ltd},
	doi = {10.1515/nanoph-2023-0598}
}

@article{inoue2006magnetophotonic,
  title={Magnetophotonic crystals},
  author={Inoue, M and Fujikawa, R and Baryshev, A and Khanikaev, A and Lim, PB and Uchida, H and Aktsipetrov, O and Fedyanin, A and Murzina, T and Granovsky, A},
  journal={Journal of Physics D: Applied Physics},
  volume={39},
  number={8},
  pages={R151-R161},
  year={2006},
  doi={10.1088/0022-3727/39/8/R01}
}

@InCollection{klos2021magnonics,
  author    = {Klos, J. and Lyubchanskii, Igor and Krawczyk, Maciej and Gruszecki, Pawel and Mieszczak, Szymon and Rychły, Justyna and Dadoenkova, Yuliya and Dadoenkova, N.},
  booktitle = {{Optomagnonic Structures: Novel Architectures for Simultaneous Control of Light and Spin Waves}},
  title     = {{Magnonics and Confinement of Light in Photonic–Magnonic Crystals}},
  year      = {2021},
  month     = {01},
  pages     = {79-134},
  doi       = {https://doi.org/10.1142/9789811220050_0002},
}

@article{levy2001flat,
  title={{Flat-top response in one-dimensional magnetic photonic bandgap structures with Faraday rotation enhancement}},
  author={Levy, M and Yang, HC and Steel, MJ and Fujita, J},
  journal={Journal of Lightwave Technology},
  volume={19},
  number={12},
  pages={1964},
  year={2001},
  publisher={OSA},
  doi={10.1109/50.971692}
}

@Article{lyubchanskii2003magnetic,
  author  = {Lyubchanskii, Igor and Dadoenkova, N. and Lyubchanskii, M and Shapovalov, E. and Rasing, Th},
  journal = {Journal of Physics D: Applied Physics},
  title   = {{Magnetic Photonic Crystals}},
  year    = {2003},
  month   = {09},
  number  = {18},
  pages   = {R277},
  volume  = {36},
  doi     = {10.1088/0022-3727/36/18/R01},
}

@article{krichevsky2024spatially,
author = {Krichevsky, Denis and Ozerov, Vladislav and Bel’kova, Alexandra and Sylgacheva, Daria and Kalish, Andrey and Evstigneeva, S. and Pakhomov, Alexander and Mikhailova, Tatyana and Lyashko, Sergey and Kudryashov, Alexander and Semuk, Evgeny and Chernov, Alexander and Vladimir, Berzhansky and Belotelov, Vladimir},
year = {2024},
month = {01},
pages = {299-306},
title = {{Spatially inhomogeneous inverse Faraday effect provides tunable nonthermal excitation of exchange dominated spin waves}},
volume = {13},
journal = {Nanophotonics},
doi = {10.1515/nanoph-2023-0626}
}

@Article{ignatyeva2024asymmetric,
  author    = {Ignatyeva, DO and Mikhailova, TV and Kapralov, PO and Lyashko, SD and Berzhansky, VN and Belotelov, VI},
  journal   = {Physical Review Applied},
  title     = {{Asymmetric Faraday effect caused by a break of spatial symmetry}},
  year      = {2024},
  number    = {4},
  pages     = {044064},
  volume    = {22},
  doi       = {https://doi.org/10.1103/PhysRevApplied.22.044064},
  publisher = {APS},
}

@Article{johnson1972optical,
  author    = {Johnson, Peter B and Christy, R. W.},
  journal   = {Physical Review B},
  title     = {Optical constants of the noble metals},
  year      = {1972},
  number    = {12},
  pages     = {4370},
  volume    = {6},
  doi       = {10.1103/PhysRevB.6.4370},
  publisher = {APS},
}

@article{Seidov2025,
  title={Dual origin of effective axion response},
  author={Seidov, Timur Z and Barredo-Alamilla, Eduardo and Bobylev, Daniel A and Shaposhnikov, Leon and Mazanov, Maxim and Gorlach, Maxim A},
  journal={Nature Communications},
  volume={16},
  number={1},
  pages={5942},
  year={2025},
  publisher={Nature Publishing Group UK London},
  doi ={10.48550/arXiv.2407.20767}
}

@Article{Nenno2020,
  author    = {Dennis M. Nenno and Christina A. C. Garcia and Johannes Gooth and Claudia Felser and Prineha Narang},
  journal   = {Nature Reviews Physics},
  title     = {Axion physics in condensed-matter systems},
  year      = {2020},
  month     = {sep},
  number    = {12},
  pages     = {682-696},
  volume    = {2},
  doi       = {10.1038/s42254-020-0240-2},
  publisher = {Springer Science and Business Media {LLC}},
}

@article{shaposhnikov2023emergent,
  title={Emergent axion response in multilayered metamaterials},
  author={Shaposhnikov, Leon and Mazanov, Maxim and Bobylev, Daniel A and Wilczek, Frank and Gorlach, Maxim A},
  journal={Physical Review B},
  volume={108},
  number={11},
  pages={115101},
  year={2023},
  publisher={APS},
  doi={10.1103/PhysRevB.108.115101}
}

@Article{wittekoek1975magneto,
  author    = {Wittekoek, S and Popma, Th JA and Robertson, JM and Bongers, PF},
  journal   = {Physical Review B},
  title     = {{Magneto-optic spectra and the dielectric tensor elements of bismuth-substituted iron garnets at photon energies between 2.2-5.2 eV}},
  year      = {1975},
  number    = {7},
  pages     = {2777},
  volume    = {12},
  doi       = {https://doi.org/10.1103/PhysRevB.12.2777},
  publisher = {APS},
}

@article{Xiao2014Apr,
	author = {Xiao, Meng and Zhang, Z. Q. and Chan, C. T.},
	title = {{Surface Impedance and Bulk Band Geometric Phases in One-Dimensional Systems}},
	journal = {Physical Review X},
	volume = {4},
	number = {2},
	pages = {021017},
	year = {2014},
	month = apr,
	publisher = {American Physical Society},
	doi = {10.1103/PhysRevX.4.021017}
}

@article{Gibbs1988Sep,
	author = {Gibbs, Doon and Harshman, D. R. and Isaacs, E. D. and McWhan, D. B. and Mills, D. and Vettier, C.},
	title = {{Polarization and Resonance Properties of Magnetic X-Ray Scattering in Holmium}},
	journal = {Physical Review Letters},
	volume = {61},
	number = {10},
	pages = {1241-1244},
	year = {1988},
	month = sep,
	publisher = {American Physical Society},
	doi = {10.1103/PhysRevLett.61.1241}
}

@article{Hepting2018Nov,
	author = {Hepting, M. and Green, R. J. and Zhong, Z. and Bluschke, M. and Suyolcu, Y. E. and Macke, S. and Frano, A. and Catalano, S. and Gibert, M. and Sutarto, R. and He, F. and Cristiani, G. and Logvenov, G. and Wang, Y. and van Aken, P. A. and Hansmann, P. and Le Tacon, M. and Triscone, J.-M. and Sawatzky, G. A. and Keimer, B. and Benckiser, E.},
	title = {{Complex magnetic order in nickelate slabs}},
	journal = {Nature Physics},
	volume = {14},
	pages = {1097-1102},
	year = {2018},
	month = nov,
	issn = {1745-2481},
	publisher = {Nature Publishing Group},
	doi = {10.1038/s41567-018-0218-5}
}

@article{Simeth2023Jun,
	author = {Simeth, Wolfgang and Bauer, Andreas and Franz, Christian and Aqeel, Aisha and Bereciartua, Pablo J. and Sears, Jennifer A. and Francoual, Sonia and Back, Christian H. and Pfleiderer, Christian},
	title = {{Resonant Elastic X-Ray Scattering of Antiferromagnetic Superstructures in ${\mathrm{EuPtSi}}_{3}$}},
	journal = {Physical Review Letters},
	volume = {130},
	number = {26},
	pages = {266701},
	year = {2023},
	month = jun,
	publisher = {American Physical Society},
	doi = {10.1103/PhysRevLett.130.266701}
}

@article{Kim2018Nov,
	author = {Kim, Min Gyu and Miao, Hu and Gao, Bin and Cheong, S.-W. and Mazzoli, C. and Barbour, A. and Hu, Wen and Wilkins, S. B. and Robinson, I. K. and Dean, M. P. M. and Kiryukhin, V.},
	title = {{Imaging antiferromagnetic antiphase domain boundaries using magnetic Bragg diffraction phase contrast}},
	journal = {Nature Communication},
	volume = {9},
	number = {5013},
	pages = {5013},
	year = {2018},
	month = nov,
	issn = {2041-1723},
	publisher = {Nature Publishing Group},
	doi = {10.1038/s41467-018-07350-3}
}

@article{Belyakov1989Aug,
	author = {Belyakov, V. A. and Dmitrienko, Vladimir E.},
	title = {{Polarization phenomena in x-ray optics}},
	journal = {Sov. Phys. Usp.},
	volume = {32},
	number = {8},
	pages = {697},
	year = {1989},
	month = aug,
	issn = {0038-5670},
	publisher = {IOP Publishing},
	doi = {10.1070/PU1989v032n08ABEH002748}
}

@Book{Joann,
  author    = {J.D. Joannopoulos and S.G. Johnson and J.N. Winn and R.D. Meade},
  publisher = {Princeton University Press, Princeton},
  title     = {{Photonic Crystals. Molding the Flow of Light}},
  year      = {2008},
}

@Article{Semnani2020,
  author    = {Semnani, Behrooz and Flannery, Jeremy and Al Maruf, Rubayet and Bajcsy, Michal},
  journal   = {Light: Science \& Applications},
  title     = {Spin-preserving chiral photonic crystal mirror},
  year      = {2020},
  issn      = {2047-7538},
  month     = Feb,
  number    = {1},
  pages     = {23},
  volume    = {9},
  doi       = {10.1038/s41377-020-0256-5},
  publisher = {Springer Science and Business Media LLC},
}

@article{kahl2003optical,
  title={{Optical transmission and Faraday rotation spectra of a bismuth iron garnet film}},
  author={Kahl, S and Popov, V and Grishin, Alexander M},
  journal={Journal of Applied Physics},
  volume={94},
  number={9},
  pages={5688-5694},
  year={2003},
  publisher={American Institute of Physics},
  doi={10.1063/1.1618935}
}

@Article{Yablonovitch1987,
  author    = {Yablonovitch, Eli},
  journal   = {Physical Review Letters},
  title     = {{Inhibited Spontaneous Emission in Solid-State Physics and Electronics}},
  year      = {1987},
  issn      = {0031-9007},
  month     = May,
  number    = {20},
  pages     = {2059-2062},
  volume    = {58},
  doi       = {10.1103/physrevlett.58.2059},
  publisher = {American Physical Society (APS)},
}

@Article{Ozawa2019,
  author    = {Ozawa, Tomoki and Price, Hannah M. and Amo, Alberto and Goldman, Nathan and Hafezi, Mohammad and Lu, Ling and Rechtsman, Mikael C. and Schuster, David and Simon, Jonathan and Zilberberg, Oded and Carusotto, Iacopo},
  journal   = {Reviews of Modern Physics},
  title     = {Topological photonics},
  year      = {2019},
  issn      = {1539-0756},
  month     = Mar,
  number    = {1},
  pages     = {015006},
  volume    = {91},
  doi       = {10.1103/revmodphys.91.015006},
  publisher = {American Physical Society (APS)},
}

@Article{Johnson2000,
  author    = {Johnson, Steven G. and Villeneuve, Pierre R. and Fan, Shanhui and Joannopoulos, J. D.},
  journal   = {Physical Review B},
  title     = {Linear waveguides in photonic-crystal slabs},
  year      = {2000},
  issn      = {1095-3795},
  month     = Sept,
  number    = {12},
  pages     = {8212-8222},
  volume    = {62},
  doi       = {10.1103/physrevb.62.8212},
  publisher = {American Physical Society (APS)},
}

@Article{Belotelov2005,
  author    = {Belotelov, V. I. and Zvezdin, A. K.},
  journal   = {Journal of the Optical Society of America B},
  title     = {Magneto-optical properties of photonic crystals},
  year      = {2005},
  issn      = {1520-8540},
  month     = Jan,
  number    = {1},
  pages     = {286},
  volume    = {22},
  doi       = {10.1364/josab.22.000286},
  publisher = {Optica Publishing Group},
}

@Article{Hughes2011,
  author    = {Hughes, Taylor L. and Prodan, Emil and Bernevig, B. Andrei},
  journal   = {Physical Review B},
  title     = {Inversion-symmetric topological insulators},
  year      = {2011},
  issn      = {1550-235X},
  month     = June,
  number    = {24},
  pages     = {245132},
  volume    = {83},
  doi       = {10.1103/physrevb.83.245132},
  publisher = {American Physical Society (APS)},
}

\end{document}


\title{Supplementary Materials.  
Magnetophotonic crystals with antiferromagnetic order}

\author{Elina A. Kokurina}
\altaffiliation{These authors contributed equally to this work}
\affiliation{School of Physics and Engineering, ITMO University, Saint Petersburg 197101, Russia}

\author{Aleksandra V. Otinova}
\altaffiliation{These authors contributed equally to this work}
\affiliation{Faculty of Physics, Lomonosov Moscow State University, Leninskie gori, Moscow 119991, Russia}

\author{Maxim~Mazanov}
\affiliation{School of Physics and Engineering, ITMO University, Saint Petersburg 197101, Russia}

\author{Vladimir I. Belotelov}
\affiliation{Faculty of Physics, Lomonosov Moscow State University, Leninskie gori, Moscow 119991, Russia}
\affiliation{Russian Quantum Center, Moscow 121205, Russia}

\author{Maxim A. Gorlach}
\email{m.gorlach@metalab.ifmo.ru}
\affiliation{School of Physics and Engineering, ITMO University, Saint Petersburg 197101, Russia}

\maketitle

\tableofcontents

\section{Eigenmodes of a gyrotropic medium}

Magnetophotonic crystal considered in the main text consists of magneto-optical layers with the out-of-plane magnetization captured by the permittivity tensor
\begin{equation}\label{eqs:Permittivity}
\hat{\eps}=
\begin{pmatrix}
\eps&&ig&&0\\-ig&&\eps&&0\\0&&0&&\eps
\end{pmatrix},
\end{equation}
where $\eps$ is a dielectric constant,  $g$ is the gyrotropy corresponding to the $z$-oriented static magnetic field and permeability $\mu$ is set to unity. In this section, for notation consistency, we present the refractive indices and the polarizations of the eigenmodes supported by a single gyrotropic layer.

We study the propagation of a monochromatic electromagnetic wave $\vc{E}=\vc{E}_0\,e^{-i\omega t + i{kz}}$ along the $z$ axis (Faraday geometry), where $\omega$ is the frequency of the wave, $k$ is the wave number, and $q=\omega/c$ is the vacuum wave number. The refractive index for the eigenmode is denoted as $n\equiv k/q$.

In the absence of external sources, Maxwell's equations for plane monochromatic waves read:
  \begin{align}
      \vc{k}\times\vc{E}=q\,\vc{B}; \qquad \vc{k} \times \vc{H}=-q\,\vc{D}.
  \end{align}

Combining the two equations above and excluding the magnetic field, we obtain:
\begin{equation}
    \mathbf{D}-n^2\mathbf{E}_{\bot}=0,
\end{equation}
where $\vc{E}_{\bot}=\vc{E}-\vc{k}\,(\vc{k}\cdot\vc{E})/k^2$ is a component of electric the field orthogonal to the wave vector, i.e. $z$ axis. Using the explicit expression for the permittivity tensor Eq.~\eqref{eqs:Permittivity}, we obtain a linear system of equations:
\begin{align}\label{eq:Eigensystem}
\begin{pmatrix}
\eps-n^2&&ig\\-ig&&\eps-n^2
\end{pmatrix}
\begin{pmatrix}
    E_x\\E_y
\end{pmatrix}
=0.
\end{align}
This matrix equation has nontrivial solutions only if its determinant is equal to zero. This yields two refractive indices of the eigenmodes: $n_{\pm}=\sqrt{\eps\mp g}$. To identify the eigenmode polarization, we return to Eq.~\eqref{eq:Eigensystem} substituting $n=n_{\pm}$. This results in $E_y=\pm iE_x$.

Hence, we obtain a well-known result: the eigenmodes of a gyrotropic medium have circular polarizations with electric field proportional to the circular vectors $\mathbf{e}_{\pm}=\mathbf{e}_x\pm i\mathbf{e}_y$. The respective refractive indices are $n_{\pm}=\sqrt{\eps\mp g}$. The difference of the eigenmode refractive indices causes the rotation of polarization plane for the linearly polarized light~-- the Faraday effect. While the physics of gyrotropic media is  well known, the phenomena in antiferromagnetic stacks of gyrotropic layers have been investigated much less~-- which is the subject of the sections below.

\section{Dispersion equation for antiferromagnetic magnetophotonic crystal}

In this section, we derive the dispersion equation for the Bloch modes in a periodic antiferromagnetic magnetophotonic crystal formed by the gyrotropic layers where the gyrotropy sign changes from layer to layer. For clarity, we consider the waves propagating along the  $z$-axis, which is orthogonal to the layers.

To compute the bulk dispersion, we employ the transfer matrix method. We introduce the transfer matrix $\hat{M}$ as $\vc{F}(z)=\hat{M}\,\vc{F}(0)$, where $\mathbf{F}=(\mathbf{E}, \mathbf{H})^T$ is a $4\times 1$ vector composed of the components of electric and magnetic fields tangential to the layers. Note that this vector is continuous at the boundaries of the layers due to the boundary conditions.

As the modes of each individual gyrotropic layer are circularly polarized, we expand the field in each layer of the structure $(\alpha=1,2)$ in the basis of left- and right circular polarizations  $E_{\alpha,\eta}$ and $H_{\alpha,\eta}$, labeled by the index $\eta=\pm 1$ indicating RCP and LCP, respectively. Each of these components, in turn, is presented as a combination of forward- and backward-propagating waves indicated by the upper index $+$ and $-$. At $z=0$ the expansion reads:
\begin{align}
\label{Match}
\begin{cases}
&E_{\alpha,\eta}(0)=E_{\alpha,\eta}^+ +E_{\alpha,\eta}^-\:,\\
&H_{\alpha,\eta}(0)=Y_{\alpha}^{\eta}E^+_{\alpha,\eta}-Y_{\alpha}^{\eta}E^-_{\alpha,\eta}.
\end{cases}
\end{align}
Here  $k_{\alpha,\eta}=qn_{\alpha}^{\eta}$, 
{$Y_{\alpha}^{\eta}=-i\eta \sqrt{(\eps_{\alpha}-\eta g)/\mu_{\alpha}}$} is the admittance for the forward-propagating wave, $n_{\alpha}^{\eta}={\sqrt{(\eps_{\alpha}-\eta g)\,\mu_{\alpha}}}$ is the refractive index,  $\mu_{\alpha}$ is the permeability of the layer and $\alpha=1,2$ denotes the layer number.

For $z=d$, the same expansion takes the form:
\begin{equation}
\left\{\begin{aligned}
    &E_{\alpha,\eta}(d)=E^+_{\alpha,\eta}e^{ik_{\alpha,\eta}d}+E^-_{\alpha,\eta}e^{-ik_{\alpha,\eta}d}\:,\\
    &H_{\alpha,\eta}(d)=Y_{\alpha}^{\eta}E^+_{\alpha,\eta}e^{ik_{\alpha,\eta}d}-Y_{\alpha}^{\eta}E^-_{\alpha,\eta}e^{-ik_{\alpha,\eta}d}.\\
    \end{aligned} \right.
\end{equation}
Eliminating $E^{\pm}_{\alpha,\eta}$ from these equations yields the transfer matrix for each of the layers $\alpha=1,2$:
\begin{equation}
\label{mmatrix}
\hat{M}^{\eta}_{\alpha}(d)=
\begin{pmatrix}    \cos(k^{\eta}_{\alpha}d)&&\frac{i\sin(k^{\eta}_{\alpha}d)}{Y_{\alpha}^{\eta}}\\
iY_{\alpha}^{\eta}\,\sin(k^{\eta}_{\alpha}d)&&\cos(k^{\eta}_{\alpha}d)
\end{pmatrix}.
\end{equation}
Note that the determinant of the transfer matrix is equal to unity. Given the transfer matrix, the fields with a given circular polarization at $z$ and $z+a$ are related as $\mathbf{F}_{\eta}(z+a)=\hat{M}^{\eta}_{2}(d)\hat{M}^{\eta}_{1}(d)\mathbf{F}_{\eta}(z)=\hat{M}^{\eta}(z)\mathbf{F}_{\eta}(z)$. On the other hand, due to the periodicity of the system, the same fields are related via $\mathbf{F}_\eta(z+a)=e^{ika}\mathbf{F}_\eta(z)$, where $k$ is the Bloch wave number and $a=2d$ is the period of the structure. Combining together the two expressions for $\mathbf{F}_\eta(z)$, we obtain the system of equations:
\begin{equation}\label{eq:eigenval1}
\left[\hat{M}^{\eta}_{2}(d)\hat{M}^{\eta}_{1}(d)-\hat{I}e^{ika}\right]\,\vc{F}_\eta(z)=0\:.
\end{equation}
This system has nontrivial solutions provided
\begin{equation}
\label{eigenvalue}
    |\hat{M}^{\eta}_{2}(d)\hat{M}^{\eta}_{1}(d)-\hat{I}e^{ika}|=0,
\end{equation}
which defines the dispersion of the bulk modes in a periodic crystal. The dispersion equations for right- and left circular polarizations decouple and take the form
\begin{equation}
\cos(ka)=F(q)\equiv\cos(n^{\eta}_1qd)\cos(n^{\eta}_2qd)-\frac{1}{2}\left(\frac{Y^{\eta}_2}{Y^{\eta}_1}+\frac{Y^{\eta}_1}{Y^{\eta}_2}\right)\sin(n^{\eta}_1qd)\sin(n^{\eta}_2qd)\:.
\label{:dispersionsameeps}
\end{equation}
If the permittivities of the layers 1 and 2 are equal, the dispersion equations for RCP and LCP coincide. Therefore, for simplicity, we drop the superscript $\eta$ in the following.

\begin{figure}
    \centering    \includegraphics[width=0.75\linewidth]{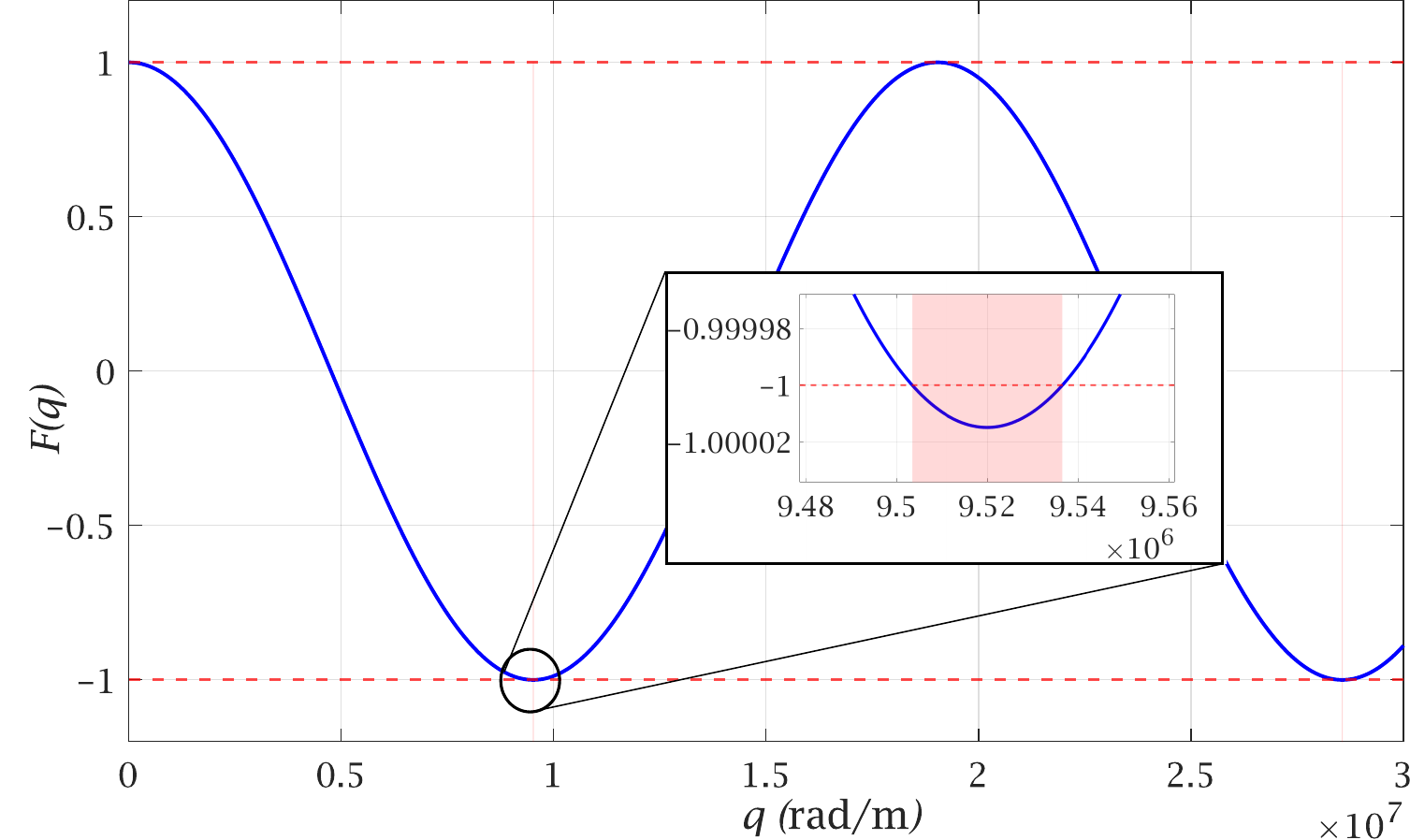}
    \caption{Right-hand side of the dispersion equation~\eqref{:dispersionsameeps} for a periodic MPC. When $|F(q)|>1$, the Bloch momentum $k$ is complex, which corresponds to the bandgap. Parameters of the simulation: $\eps=5.5$, $d=70$ nm, $g=0.03$. The first bandgap is shaded.}
    \label{fig:illustration}
\end{figure}

We aim to find the frequencies corresponding to the bulk bandgap of a periodic crystal. To that end, in Fig.~\ref{fig:illustration} we plot the right-hand side of the dispersion equation~\eqref{:dispersionsameeps}. If $-1<F(q)<1$, Eq.~\eqref{:dispersionsameeps} has real-valued solutions for $k$, which form the photonic band. On the contrary, once $|F(q)|>1$, the wave number becomes complex which physically describes the attenuation of the waves. The respective frequencies are associated with the bulk bandgap. Note that the center-of-bandgap frequency corresponds to the extremum of the function $F(q)$ and hence satisfies the condition
\begin{equation}\label{eq:center-of-bandgap}
\df{F}{q}=0\:.
\end{equation}
For further analysis, we present the dispersion equation as
\begin{equation}\label{eq:disp2}
\cos ka=\cos\left(q(n_1+n_2)d\right)+\left[1-\frac{1}{2}\left(\frac{Y_2}{Y_1}+\frac{Y_1}{Y_2}\right)\right]\,\sin(qn_1d)\,\sin(qn_2 d)\:.
\end{equation}

We assume that the gyrotropy $g$ is quite weak, i.e. $|g|\ll\eps$, while $\mu=1$. Then the term
\begin{equation*}
1-\frac{1}{2}\left(\frac{Y_2}{Y_1}+\frac{Y_1}{Y_2}\right)\approx -\frac{g^2}{2\eps^2}
\end{equation*}
and can be neglected, while
\begin{equation*}
F(q)\approx \cos(n_1qd+n_2 qd)\:.
\end{equation*}
In this simplified description local extrema of $F(q)$ and the respective center-of-bandgap frequencies are given by the expression 
\begin{equation}\label{eq:gapposition1}
\omega_m=\frac{\pi c m}{(n_1+n_2)d}\:,
\end{equation}
where $m=1,2,\dots$, which for the lowest gap can be simplified to yield
\begin{equation}\label{eq:gapposition}
\omega_B \simeq \frac{\pi c}{2\sqrt{\eps}\,d}
\end{equation}
in agreement with the main text.

Another relevant parameter is the width of the photonic gap $\Delta \omega$. From Fig.~\ref{fig:illustration} it is seen that the width of the gap is given by the frequency interval where the function $|F(q)|$ exceeds unity. We introduce the index detuning between the two layers $\Delta n=\frac{n_2-n_1}{2}$ and the average refractive index $n=\frac{n_2+n_1}{2}$. We also use the expansion:
\begin{align}
1-\frac{1}{2}\left[\frac{Y_1}{Y_2}+\frac{Y_2}{Y_1}\right]\approx-2\,\left(\frac{\Delta n}{n}\right)^2\:.
\end{align}
At the band edge,  $\cos(ka)=-1$, while the edge-of-bandgap frequency is given by $\om_B+\Delta\om/2$. The dispersion equation~\eqref{eq:disp2} then yields
\begin{equation*}
-1=\cos\left(\pi+\Delta q n d\right)-2\,\left(\frac{\Delta n}{n}\right)^2\,\sin^2\left(q_B\,n d\right)\:.
\end{equation*}
Expanding the right-hand side and keeping the terms up to $\Delta n^2$ and $\Delta q^2$, we recover
\begin{equation}
\Delta q=\frac{2\Delta n}{n^2\,d}\:.
\end{equation}
Now we recall that $n\approx \sqrt{\eps}$ and $\Delta n=|g|/(2\sqrt{\eps})$. This finally yields the expression for the width of photonic gap
\begin{equation}
    \Delta \omega=\frac{c|g|}{d\, \varepsilon^{3/2}}
\end{equation}
in agreement with the main text.

Next we briefly comment on the properties of the structure with non-zero contrast $\eps_1\neq\eps_2$. The dispersion laws for RCP and LCP modes no longer coincide. Since $n_{\alpha}^{\eta}=\sqrt{\eps_{\alpha}-\eta g}$, this yields the set of the refractive indices:
\begin{align}
\begin{split}
\label{index}
&n_1^+=\sqrt{\eps_1-g},\qquad n_2^+=\sqrt{\eps_2+g};\\
&n_1^-=\sqrt{\eps_1+g},\qquad n_2^-=\sqrt{\eps_2-g}.
\end{split}
\end{align}
When $n_2^{\eta}=n_1^{\eta}$, the respective mode experiences no contrast between the two layers, and therefore the band gap closes for this polarization. This yields the conditions of gap closing $g=\frac{\varepsilon_1-\varepsilon_2}{2}$ for RCP and $g=\frac{\varepsilon_2-\varepsilon_1}{2}$ for LCP.

Besides the mode dispersion an important information is carried by the Bloch impedance of the wave. The transfer matrix for a single unit cell reads 
\begin{align}
    \hat{M^{\eta}}=\hat{M^{\eta}_2}\hat{M_1^{\eta}}=
    \begin{pmatrix}
        A_{\eta}&&B_{\eta}\\C_{\eta}&&D_{\eta}
    \end{pmatrix}.
\end{align}
The elements of this matrix are given by the expressions:
\begin{align}
&\label{matrixelements}{A_\eta}=\cos\delta^{\eta}_1\cos\delta^{\eta}_2-\frac{Y^{\eta}_1}{Y^{\eta}_2}\,\sin\delta^{\eta}_1\sin\delta^{\eta}_2; \qquad B_{\eta}=i\left(\frac{\cos\delta^{\eta}_2\,\sin\delta^{\eta}_1}{Y_1^{\eta}}+\frac{\cos\delta^{\eta}_1\,\sin\delta^{\eta}_2}{Y^{\eta}_2}\right);\\
\label{matrixelements2}
&C_{\eta}=i\left(Y_2^{\eta}\cos\delta^{\eta}_1\sin\delta^{\eta}_2+Y_1^{\eta}\cos\delta^{\eta}_2\sin\delta^{\eta}_1\right); 
\qquad D_{\eta}=\cos\delta^{\eta}_1\,\cos\delta^{\eta}_2-\frac{Y^{\eta}_2}{Y^{\eta}_1}\sin\delta^{\eta}_1\,\sin\delta^{\eta}_2,
   \end{align}
where in the case of non-magnetic layers ($\mu=1$) $Y^{\eta}_\alpha=-i\eta\,n^{\eta}_\alpha$ and $\delta_{\alpha}^\eta=qn_{\alpha}^{\eta}d$ for each layer $\alpha=1,2$. Note that all elements of the transfer matrix are real.

Combining together Eq.~\eqref{eigenvalue} and Eq.~(\ref{matrixelements}-\ref{matrixelements2}) and taking into account unity determinant of the transfer matrix, we obtain the dispersion equation in the form $\cos(ka)=\frac{A_{\eta}+D_{\eta}}{2}$. In addition to that, we also find the relation between the components of electric and magnetic field: 
\begin{equation}
    (A_{\eta}-e^{ika})E_\eta+B_{\eta}H_\eta=0.
\end{equation}
This defines the Bloch impedance of the structure:
\begin{equation}
\label{BraggSuppl}
Z^{\eta}_B=\frac{E_\eta}{H_\eta}=\frac{B_{\eta}}{e^{ika}-A_{\eta}}.
\end{equation}

Next it is instructive to consider the Bloch impedance behavior for the two limiting cases discussed in the main text: (i)~the effective-medium long-wavelength limit, and (ii)~center-of-bandgap frequency. 

In case~(i), to the leading order, 
$$
B_\eta \simeq - 2 \eta (q d) + O((q d)^3), 
$$
$$
A_\eta \simeq 1 - \frac{3(n_1^\eta)^2 + (n_2^\eta)^2}{2} (q d)^2 + O((q d)^4), 
$$
$$
D_\eta \simeq 1 - \frac{(n_1^\eta)^2 + 3(n_2^\eta)^2}{2} (q d)^2 + O((q d)^4), 
$$ 
while the Bloch eigenvalue is 
$ e^{i k a} \simeq 1 + 2 i \sqrt{\varepsilon}\, qd - 2 \varepsilon\, (qd)^2 $. 
At the same time, $ (n_1^\eta)^2 + (n_2^\eta)^2 = 2 \varepsilon$. This allows to simplify the Bloch impedance as 
\begin{equation} \label{Z_LWL}
    Z^{\eta}_{B,  \,  q d \ll 1}
    \simeq
    \frac{ -2 \eta (q d)}{2 i \sqrt{\varepsilon} (qd) - \eta g (qd)^2} + O((qd)^2) 
    \simeq 
    \frac{i \eta}{\sqrt{\varepsilon}} + \frac{g}{2 \varepsilon} (qd) + O((qd)^2) . 
\end{equation}
Here, the first term $\frac{i \eta}{\sqrt{\varepsilon}}$ is the homogeneous medium first-order approximation with $i$ stemming from the circular basis choice, and the correction $\frac{g}{2 \varepsilon} (qd) = \chi / \varepsilon$, where $\chi=\frac{\pi}{2}g \frac{a}{\lambda} = q g d / 2$ can be interpreted as the effective axion-field correction and depends on the choice of the unit cell, similar to the effective axion response~\cite{shaposhnikov2023emergent}. 

We next consider the case (ii). At the frequency $\omega_B$ corresponding to the center of the first bandgap, the Bloch wavenumber has the form $k=\pi/a+i\kappa$ with real positive $\kappa$ which captures the attenuation of the waves inside the structure, both for RCP and LCP polarized waves: $e^{ik_\eta a}=\lambda_\eta=-e^{-\kappa a}$. 
The center-of-bandgap frequency is defined via the Eqs.~\eqref{eq:gapposition1}-\eqref{eq:gapposition}, while the relation for $\delta_\alpha^\eta$ at the bandgap center reads 
%
\[
\delta_1^\eta+\delta_2^\eta=\pi + O(g^4). 
\]
We introduce
\[
\bar n=\frac{n_1^\eta+n_2^\eta}{2}, \qquad 
\mu_\eta=\frac{n_2^\eta-n_1^\eta}{n_2^\eta+n_1^\eta},
\]
so that
$
n_{1,2}^\eta=\bar n(1\mp\mu_\eta), 
\bar n = \pi / (2 q d), 
\delta_{1,2}^\eta=\frac{\pi}{2}(1\mp\mu_\eta), 
$
and the shorthand notations
\[
c_\eta=\cos\delta_1^\eta=-\cos\delta_2^\eta=\sin\frac{\pi\mu_\eta}{2},\qquad
s_\eta=\sin\delta_1^\eta=\sin\delta_2=\cos\frac{\pi\mu_\eta}{2}.
\] 
For RCP/LCP polarizations, 
\[
\mu_+=+\mu,\qquad \mu_-=-\mu,\qquad
\mu=\frac{\sqrt{\varepsilon+g}-\sqrt{\varepsilon-g}}
{\sqrt{\varepsilon+g}+\sqrt{\varepsilon-g}}
\simeq \frac{g}{2\varepsilon}.
\]
With these notations, the unit-cell matrix elements read, at the gap center,
\[
A_\eta=-c_\eta^2-\frac{1-\mu_\eta}{1+\mu_\eta}s_\eta^2,\qquad
D_\eta=-c_\eta^2-\frac{1+\mu_\eta}{1-\mu_\eta}s_\eta^2,\qquad
B_\eta=\eta\,\frac{2\mu_\eta c_\eta s_\eta}{\bar n(1-\mu_\eta^2)},\qquad
C_\eta=2\eta\,\bar n\mu_\eta c_\eta s_\eta .
\]
Expansion of the impedance-relevant quantities in small $\mu$ gives
\begin{equation} \label{AB_expansion}
B_{\pm} = \pm \frac{\pi \mu^2}{\bar{n}}\left[1+\left(1-\frac{\pi^2}{6}\right) \mu^2+O\left(\mu^4\right)\right]
, \qquad 
A_{\pm}=-1 \pm 2 \mu - 2 \mu^2 \pm \left(2-\frac{\pi}{2}\right) \mu^3+O\left(\mu^4\right)
\end{equation}
The dispersion relation gives
\[
\lambda_\eta+\lambda_\eta^{-1}=A_\eta+D_\eta,
\qquad
\cosh(\kappa a)=-\frac{A_\eta+D_\eta}{2}
=
1+\frac{2\mu_\eta^2}{1-\mu_\eta^2}
\cos^2\frac{\pi\mu_\eta}{2},
\]
which, expanding in $\mu$ to leading order, gives
\[
\cosh (\kappa a) =1+2 \mu_\eta^2+\left(2-\frac{\pi^2}{2}\right) \mu_\eta^4+O\left(\mu_\eta^6\right). 
\]
Expanding the left-hand side (assuming small attenuation $\kappa a \ll 1$ for $g \ll 1$) we obtain 
\[
\cosh (\kappa a) =1+\frac{(\kappa a)^2}{2}+\frac{(\kappa a)^4}{24}+O\left((\kappa a)^6\right). 
\]
Carefully comparing the leading-order terms, we obtain 
\begin{equation} \label{kappa__a}
\kappa a = 2\mu + \left( \frac{2}{3} - \frac{\pi^2}{4} \right) \mu^3 + O(\mu^5), 
\end{equation}
thus
\begin{equation} \label{lam_expansion}
\lambda_\eta \equiv \lambda=-e^{-\kappa a}=-1+2 \mu-2 \mu^2+\left(2-\frac{\pi^2}{4}\right) \mu^3+O\left(\mu^4\right)
.
\end{equation}
Using the obtained expansions~\eqref{AB_expansion} and~\eqref{lam_expansion}, we expand the Bloch impedance~\eqref{BraggSuppl} at the center of the first bandgap which gives, for RCP polarization, 
\begin{equation} \label{Z_+_CB}
Z_{B}^{+}(\omega_B)
=
\frac{4}{\pi\bar n\,\mu}
\left[
1+\left(1-\frac{\pi^2}{48}\right)\mu^2+O(\mu^4)
\right],
\end{equation}
whereas for LCP polarization, 
\begin{equation} \label{Z_-_CB}
Z_{B}^{-}(\omega_B)
=
-\frac{\pi\mu}{4\bar n}
\left[
1+\frac{\pi^2}{48}\mu^2+O(\mu^4)
\right].
\end{equation}
Thus, at the first-gap center, the RCP/LCP impedances feature markedly distinct scaling with $\mu \simeq g / (2 \varepsilon)$: 
\[
|Z_{B,+}|\sim \mu^{-1}\gg1,\qquad
|Z_{B,-}|\sim\mu\ll1.
\]
The purely real values of the impedances are connected with the circular basis notation.

\section{Reflection and transmission coefficients}
\subsection{Reflection from the semi-infinite medium}
\label{sec3}

To describe the reflection of light from the semi-infinite photonic crystal, we match the fields at the boundary. For a circularly polarized mode $\eta$ incident from  vacuum onto the crystal surface at $z=0$, the tangential components are related as $\mathbf{E}^{\eta}_t(z)=\mathbf{E}^{\eta}_{t}(z')$ and $\mathbf{H}^{\eta}_t(z)=\mathbf{H}^{\eta}_{t}(z')$, where $z$ denotes the vacuum side and $z'$ denotes the photonic crystal side. These continuity relations relate the fields incident at the interface (comprising the incident and reflected fields) to the transmitted fields: 
\begin{align}
\label{frensel}
&E^{\eta}_{in}+E^{\eta}_{r}=E^{\eta}_{tr},\\
&H^{\eta}_{in}+H^{\eta}_{r}=H^{\eta}_{tr}.
\end{align}
With $H^{\eta}_{in}=Y^{\eta}_0E^{\eta}_{in}$, $H^{\eta}_{r}=-Y^{\eta}_0\,E^{\eta}_{r}$ and $H^{\eta}_{tr}=Y^{\eta}_BE^{\eta}_{tr}$, where $Y^{\eta}_0=1/Z_0^{\eta} = - i \eta$ is the vacuum admittance (where the vacuum impedance is $Z^{\eta}_0 = i \eta$) and $Y_B^\eta=1/Z_B^\eta$ is the Bloch wave admittance. Introducing the reflection coefficient for the corresponding polarization $r_{\eta}=E_{r}/E_{in}$ and applying the boundary conditions yields the Frensel formulas:
\begin{equation}
\label{frensel_r}
r_{\eta}=\frac{Z^{\eta}_B-Z^{\eta}_0}{Z^{\eta}_B+Z^{\eta}_0}.
\end{equation}
Here, we can use the approximated expressions for impedance derived above for the two limiting cases: metamaterial limit [Eq.~\eqref{Z_LWL}] and center-of-bandgap frequency [Eqs.~\eqref{Z_+_CB}-\eqref{Z_-_CB}]. Then, we can further transform the reflection coefficients to the linear (Cartesian) basis:
\begin{equation}
\label{Cartesian}
r_{xx}=\frac{1}{2}\left(r_++r_-\right),\qquad r_{xy}=\frac{1}{2i}\left(r_+-r_-\right)
. 
\end{equation}

We start from the long-wavelength case. Substituting the long-wavelength limit result Eq.~\eqref{Z_LWL} into Eq.~\eqref{frensel_r}, we obtain 
\[
r_\eta^{q d \ll 1}
\simeq
r_0
-
i\eta\,\frac{g}{(1+n)^2} qd
=
r_0
-
i\eta\,\frac{2 \chi}{(1+n)^2} 
,
\]
where $ n=\sqrt{\varepsilon} $ and 
$
r_0=\frac{1-n}{1+n}. 
$
The Cartesian components of the reflection amplitudes read 
\begin{equation} \label{rxx_LWL}
r_{xx}^{q d \ll 1}
=
\frac{1-\sqrt{\varepsilon}}{1+\sqrt{\varepsilon}}
+O((qd)^2),
\end{equation}
\begin{equation} \label{rxy_LWL}
r_{xy}^{q d \ll 1}
=
-\frac{2 \chi}{(1+\sqrt{\varepsilon})^2}
+O((qd)^2),
\end{equation}
which matches the result in Ref.~\cite{shaposhnikov2023emergent}.

\begin{figure}
    \centering
    \includegraphics[width=0.40\linewidth]{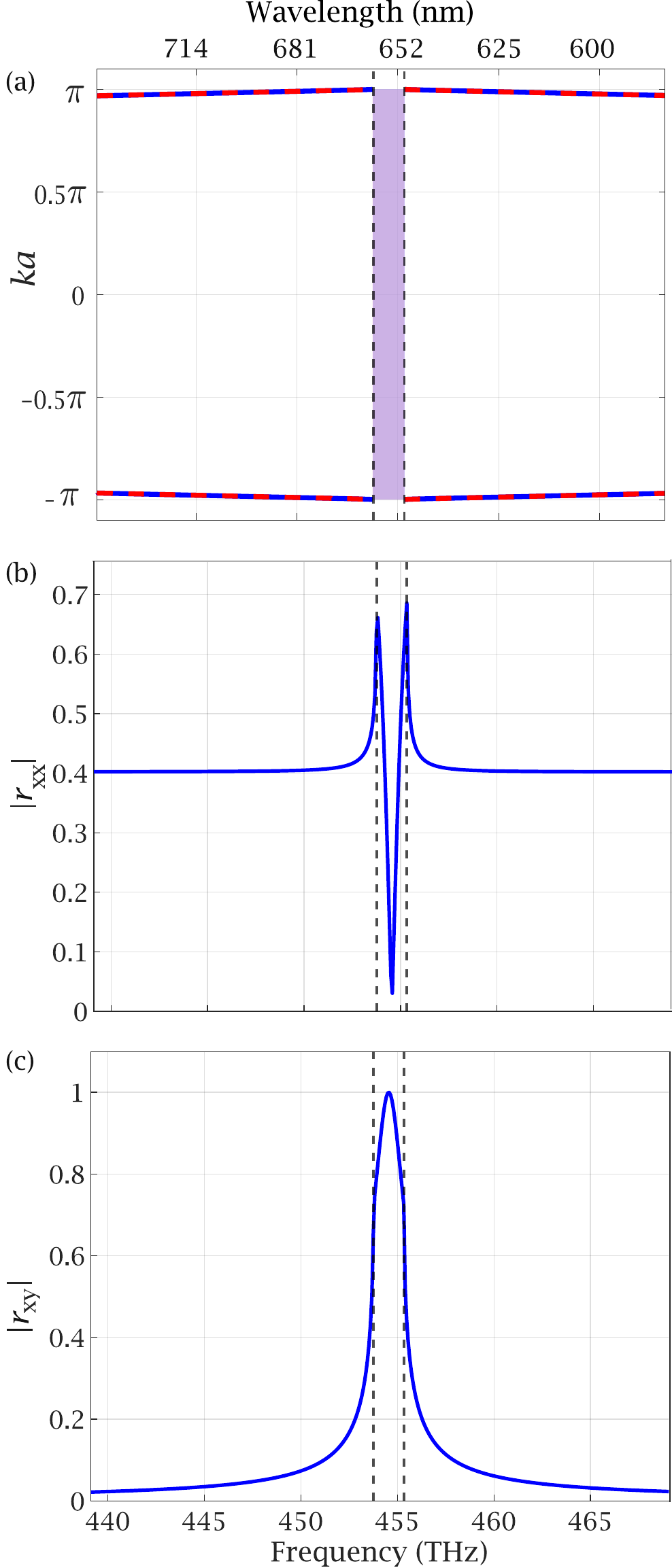}
    \caption{Photonic band gap and reflection properties of the semi-infinite magneto-photonic crystal with $\eps_1=\eps_2=5.5$, $g=0.03$, $d=70$~nm. (a) The dispersion of the Bloch modes close to photonic bandgap of MPC. (b,c) Co- and cross-polarized reflection coefficients close to the bandgap frequencies. A sharp maximum of $r_{xy}$ is observed, with almost unity cross-polarized reflection at the bandgap center.}
    \label{fig:axmet}
\end{figure}

Next, we proceed examining center-of-bandgap frequency. The Bloch impedances from Eqs.~\eqref{Z_+_CB}-\eqref{Z_-_CB} read, to leading order,
\begin{equation} \label{impedances_midgap_AB}
Z_{B}^+(\omega_B)
=
\frac{4}{\pi\bar n\mu}
\left[1+O(\mu^2)\right] \gg 1,
\qquad
Z_{B}^-(\omega_B)
=
-\frac{\pi\mu}{4\bar n}
\left[1+O(\mu^2)\right] \ll 1.
\end{equation}
Therefore for RCP, 
\[
r_+(\omega_B)
=
\frac{Z_{B}^+-i}{Z_{B}^++i}
=
1-\frac{2i}{Z_{B}^+}+O((Z_{B}^+)^{-2})
=
1-i\frac{\pi\bar n}{2}\mu+O(\mu^2),
\]
whereas for LCP, 
\[
r_-(\omega_B)
=
\frac{Z_{B}^-+i}{Z_{B}^- -i}
=
-1+2iZ_{B}^-+O((Z_{B}^-)^2)
=
-1-i\frac{\pi}{2\bar n}\mu+O(\mu^2).
\]
In Cartesian components, we thus obtain 
\begin{equation} \label{rxx_CB}
r_{xx}(\omega_B)
=
-i\frac{\pi\mu}{4}
\left(
\bar n+\frac1{\bar n}
\right)
+O(\mu^2)
=
-i g 
\frac{\pi (\varepsilon + 1)}{8 \varepsilon^{3/2}}
+O(g^2),
\end{equation}
and
\begin{equation} \label{rxy_CB}
r_{xy}(\omega_B)
=
-i
-
\frac{\pi\mu}{4}
\left(
\bar n-\frac1{\bar n}
\right)
+O(\mu^2)
=
-i - 
g 
\frac{\pi (\varepsilon - 1)}{8 \varepsilon^{3/2}}
+O(g^2).
\end{equation}
Thus \(r_{xx}(\omega_B)=O(\mu)\), whereas \(r_{xy}(\omega_B)=-i+O(\mu)\).
The corresponding polarization rotation angle is then, to leading order in $g$, 
\begin{equation}
\label{pur}
    \theta = \frac{\pi}{2} + g 
\frac{\pi (\varepsilon + 1)}{8 \varepsilon^{3/2}} + O(g^2)
    . 
\end{equation}
This signifies an important result: 
\textbf{the reflection from the semi-infinite MPC at center-of-bandgap frequency is almost purely cross-polarized to the leading order in $g$}, with small deviations proportional to $g$.
In the other words, the reflected wave is fully cross-polarized with respect to the incident polarization, and the structure acts as a perfect polarization rotator. 
As shown in Fig.~\eqref{fig:axmet}, both the position and the width of this cross-polarized resonance coincide exactly with those of the band gap, and we observe a single sharp peak within the bandgap.

{
It should be noted that the formula~\eqref{pur} is only applicable in the case of $g\neq0$, i.e., when the photonic gap has a finite width $\Delta\omega\propto g$ and the decay length $1/ \kappa$ is finite. Indeed, the derivation of this formula itself supposes the existence of $Z_B^{\eta}(\omega_B)$ which becomes zero or non-defined when the gap closes ($Z^{+}_B(\omega_B) \rightarrow \infty$ and $Z_B^-(\omega_B) \rightarrow 0$). 

The limits $g\rightarrow 0$ and $N\rightarrow\infty$ do not commute because the penetration length diverges with the vanishing gyrotropy, allowing the wave propagate without decay. For the infinite crystal, this raises a paradox since there is no second boundary and the wave cannot be propagating (it must decay to have total reflection which is the only possible outcome for a semi-infinite case).  To physically correctly evaluate the case of non-magnetic medium, we first have to set $g\rightarrow 0$  and obtain the non-magnetic crystal, closing the Bragg gap, and then make the slab infinitely large. Therefore, the above formula is only valid for $g\neq0$ and  cannot be extrapolated to the case of $g=0$, where the band gap closes and the Bragg reflection mechanism vanishes. 
}

\clearpage

\subsection{Reflection and transmission coefficient for the finite slab}

In this section, we compute the reflection and transmission coefficients for a finite stack of gyrotropic layers with the integer number of unit cells. Now, it is convenient to switch from the  basis of $(\mathbf{E},\mathbf{H})$ used in the previous section to the circular basis of $(E^+_{\rm RCP}, E^-_{\rm RCP}, E^+_{\rm LCP}, E^-_{\rm LCP}$), where the top sign means the propagating wave traveling along ($+$) and opposite to ($-$) the direction of the $z$ axis. The advantage of this representation is that, in the new basis, propagation through a homogeneous layer is expressed by  simple multiplication of the amplitudes by a phase factor, while propagation across the interface is determined by the nontrivial boundary matrix. 

At the interface between  media $i$ and $j$, the tangential components of the total fields $\mathbf{E}$ and $\mathbf{H}$ are continuous. In the above basis, this requirement is satisfied by introducing the boundary matrix $\hat{M}_{ij}$ which reads~\cite{shaposhnikov2023emergent}: 
\begin{equation}
\hat{M}_{ij}
=\frac{1}{2}
\begin{pmatrix}
1+\eta_{ij}^{\pm}&&1-\eta_{ij}^{\pm}&&0&&0\\1-\eta_{ij}^{\pm}&&1+\eta_{ij}^{\pm}&&0&&0\\
0&&0&&1+\eta_{ij}^{\mp}&&1-\eta_{ij}^{\mp}\\0&&0&&1-\eta_{ij}^{\mp}&&1+\eta_{ij}^{\mp}
\end{pmatrix}
\end{equation}
where the ratio of the refractive indices $\eta^{\pm}_{ij}=n^{\pm}_i/n^{\pm}_j$ is defined from Eq.\eqref{index}. In the case of propagation from air to the medium we set $n^{\pm}_{ij}$ =1 for the air/vacuum medium. 
The propagation matrix  through a layer $\alpha=i,j$ of thickness $d$ is diagonal in the circular basis $\hat{M}_{\alpha}(d)=\text{diag}(e^{ik_{\alpha}^+d},e^{-ik_{\alpha}^+d},e^{ik_{\alpha}^-d},e^{-ik_{\alpha}^-d})$.
The upper-left $2 \times 2$ block $\hat{M}_{RCP}$ contains the terms related to the RCP wave, and the lower-right 
$2 \times 2$ block $\hat{M}_{LCP}$  contains those related to the LCP wave. For the case of zero contrast $\eps_1=\eps_2$, the dispersions for both polarizations coincide, as do their band gaps. For nonzero contrast $\eps_1\neq\eps_2$, the total band gap corresponds to the intersection of the individual band gaps. The total transfer matrix allows us to evaluate the spectrum for each case. In a lossless system,  $\det{M}=\det{M^{LCP}}\cdot\det{M^{RCP}}=1$. From the symmetry,  $\det{M^{LCP}}=\det{M^{RCP}}=1$.

The total transfer matrix for $N$ periods (with two layers with opposite-sign $g$ within each cell) is:
\begin{equation}
\label{total}\hat{M}_{total}=\hat{M}_{2-air}\hat{M}_{1-2}(\hat{M}_{2-1}\hat{M}_{2}\hat{M}_{1-2}\hat{M}_{1})^N\hat{M}_{air-1}
\end{equation}
In terms of reflected and transmitted waves, the total transfer matrix relates the electric fields from both sides of the boundary  as:
\begin{equation}
\begin{pmatrix}
E^t_{\eta}\\0
 \end{pmatrix}
    =\hat{M}_{total,\eta}
    \begin{pmatrix}
        E^i_{\eta}\\E^r_{\eta}
    \end{pmatrix}
\end{equation}
which gives the 
reflected and transmitted fields in terms of the coefficients of the total transfer matrix:
\begin{align}
\label{rt}
    &r^{\eta}=-\frac{M^{\eta}_{21}}{M^{\eta}_{22}} \\
    &t^{\eta}=\frac{\text{det}(M^{\eta})}{M^{\eta}_{22}}=\frac{1}{M^{\eta}_{22}}.
\end{align}
To find co- and cross-polarized components, we switch back to the Cartesian basis via the transformation $E_x=\frac{E_++E_-}{2}, E_y=\frac{E_+-E_-}{2i}$.
Applying the linear transformation to our solution for the reflection coefficients yields the expressions for the finite slab:
\begin{align}
    &r_{xx}=\frac{1}{2}\left(\frac{M_{21}}{M_{22}}+\frac{M_{43}}{M_{44}}\right), \qquad r_{xy}=\frac{1}{2i}\left(\frac{M_{21}}{M_{22}}-\frac{M_{43}}{M_{44}}\right);\\
      &t_{xx}=\frac{1}{2}\left(\frac{1}{M_{22}}+\frac{1}{M_{44}}\right), \qquad t_{xy}=\frac{1}{2i}\left(\frac{1}{M_{22}}-\frac{1}{M_{44}}\right).
\end{align}
\begin{figure}
    \centering
    \includegraphics[width=0.4\linewidth]{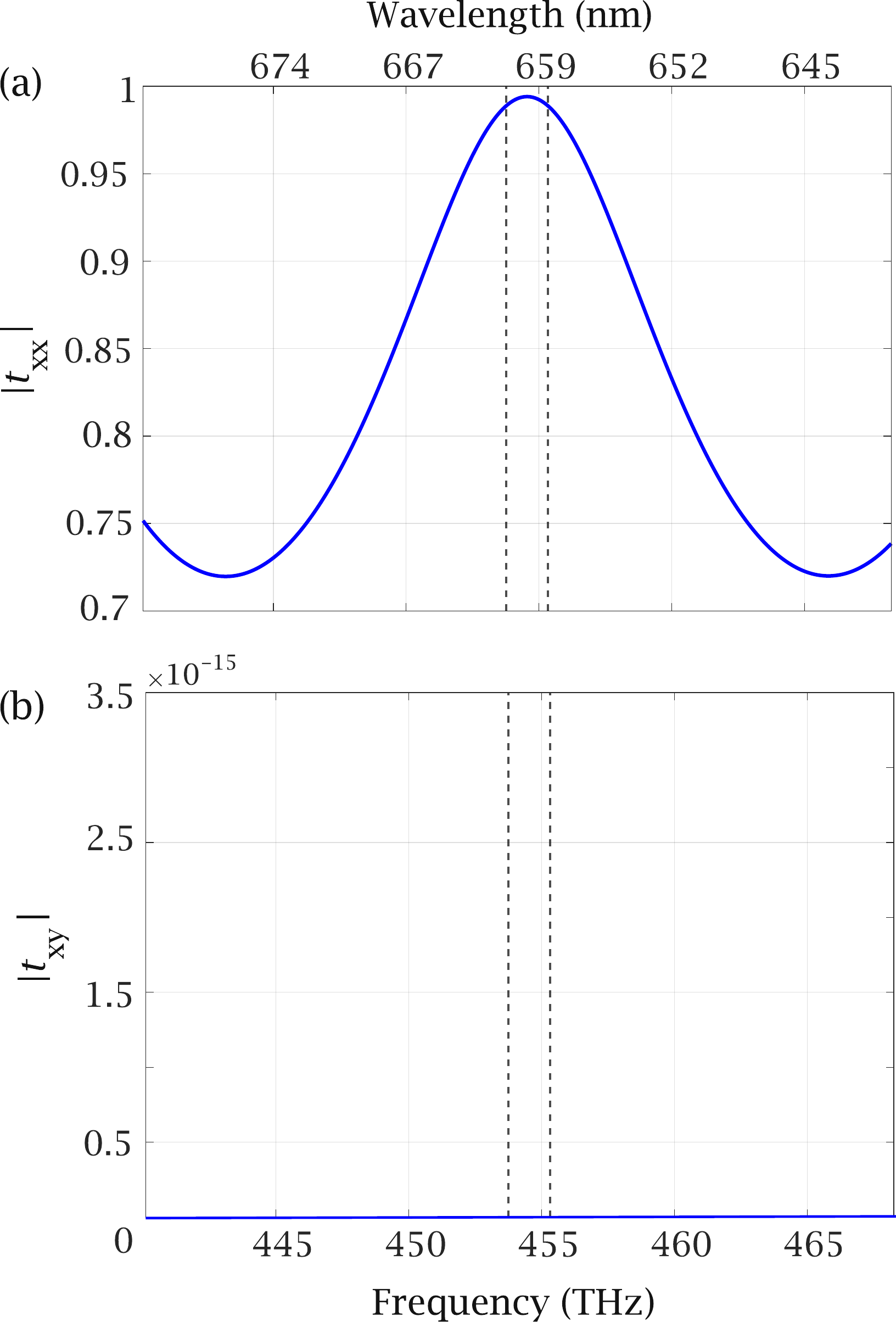}
    \caption{Co- and cross-polarized transmission coefficients for a magnetophotonic crystal. 
    Parameters of simulation: $N=20$, $d=70$~nm,  $\eps_1=\eps_2=5.5$, $g=0.03$. Cross-polarized transmission $t_{xy}$ vanishes identically due to the $\mathcal{PT}$ symmetry of the structure.}
    \label{fig:limitpdf1}
\end{figure}

Fig.~\ref{fig:limitpdf1} shows that the transmission spectra for the finite magnetophotonic crystal in addition to the reflection coefficients in Fig.~3 in the main text. We observe that the cross-polarized transmission vanishes identically due to the $\mathcal{PT}$ symmetry of the structure, while the co-polarized transmission is of the order of $1$. This large transmission is due to the narrow bandgap and insufficient thickness of the MPC, which is not enough to suppress the transmission.

\begin{figure}
    \centering
    \includegraphics[width=0.45\linewidth]{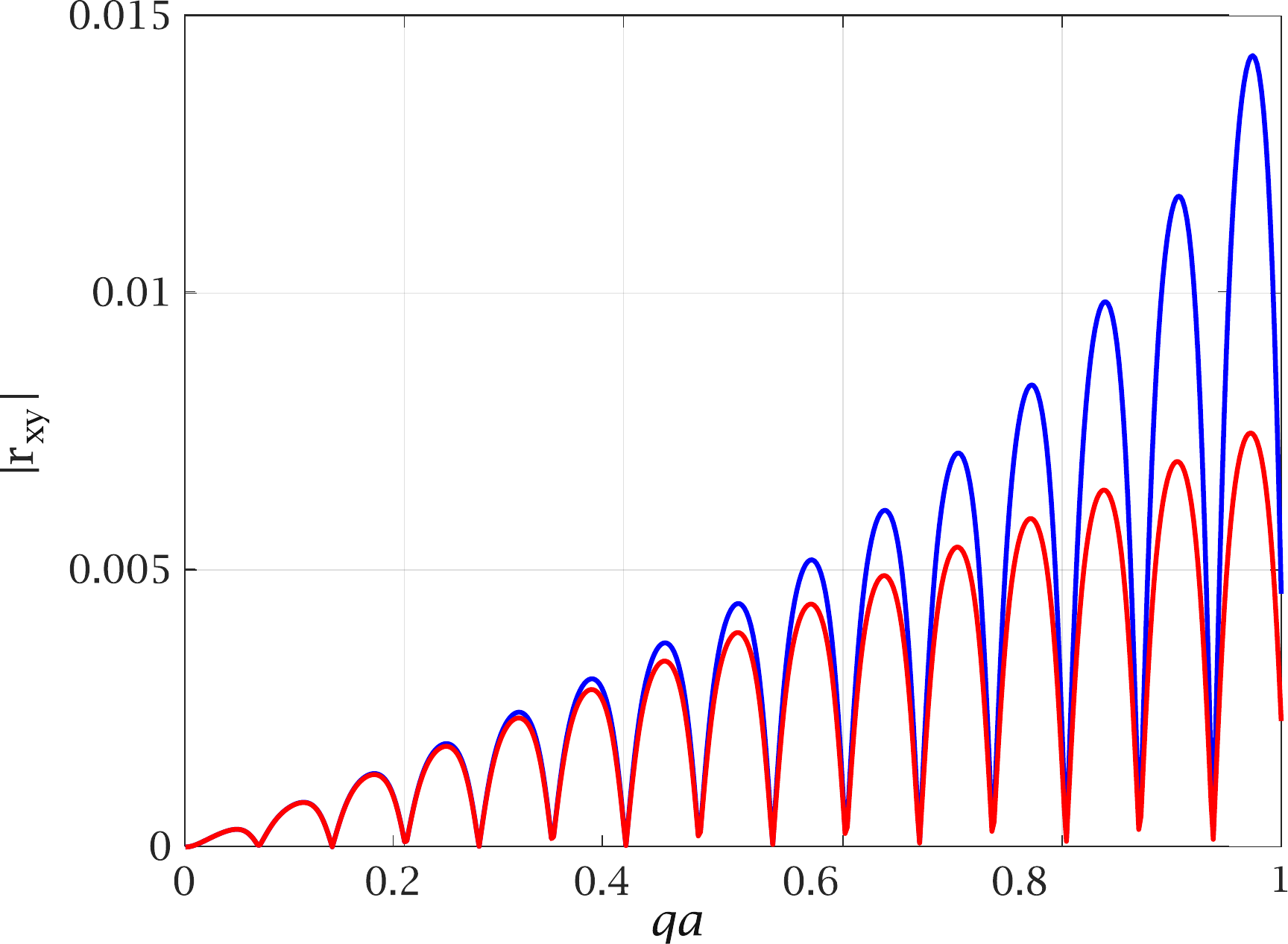}
    \caption{Cross-polarized reflection from MPC consisting of $N=20$ unit cells, $\eps=5.5$, $g=0.03$ in the metamaterial regime $qa\ll 1$. Blue line shows the result of the transfer matrix method, while the red line is the result of effective description in terms of effective permittivity and emergent axion response $\chi$.}
    \label{fig:limitpdf}
\end{figure}

Now we consider the subwavelength limit of the multilayer, where the incident wavelength covers many periods of the structure, $a\ll\lambda$, or, in other words, $qa\ll1$. In this regime, the medium can be treated as an effective medium~-- a homogeneous material characterized by effective parameters, such as the effective dielectric permittivity $\varepsilon_{eff}$, effective magnetic permeability $\mu_{eff}$, and additional bianisotropic responses such as chirality or Tellegen (axion) response. The effective medium description dramatically simplifies the analysis, reproducing the macroscopic optical response of the structure in terms of bulk constitutive relations. 
In the case of a finite slab consisting of $N$ unit cells with $\eps_1=\eps_2=\eps$, placed in the vacuum, Ref.~\cite{shaposhnikov2023emergent} has shown that such slab can be described by the equations of axion electrodynamics, which predict the reflection coefficients as 
\begin{align}
& r_{xx}=r_{yy}=\frac{(1-\eps - \chi^2)\sin L}{\Delta}, \qquad r_{xy}=-r_{yx}=\frac{2\chi \sin L}{\Delta};\\
&t_{xx}=t_{yy}=\frac{2i\sqrt{\eps}}{\Delta},\qquad t_{xy}=t_{yx}=0;\qquad L=\sqrt{\eps} N qa;\\
&\Delta=(\chi^2+1+\eps)\sin L +2i\sqrt{\eps} \cos L;
\end{align}
As shown in Fig.~\ref{fig:limitpdf}, the effective medium captures the behavior of the periodic structure well as long as the subwavelength regime holds.

\section{Topological localized states}
 
Each unit cell of the magnetophotonic crystal can be chosen with an inversion center either in the middle of layer $1$ or in the middle of layer $2$. This choice is known as the inversion-symmetric unit cell, and the two choices are related by a shift of the origin by half a period and are generally not equivalent, and they can possess different topological invariants quantized by the inversion symmetry. 
According to the bulk-boundary correspondence, 
the localized state 
is guaranteed to emerge 
at the interface between the two photonic crystals with different topological invariants, with both having terminations consistent with their respective inversion-symmetric unit cells (i.e. without cutting these unit cells by the boundary). It should be noted that in the presence of $\mathcal{PT}$-symmetry the field distribution for both RCP and LCP polarizations coincide, so both circular polarizations share the same band gap and, as we show below, produce degenerate localized states with identical field distributions with the frequency exactly in the center of the bandgap. To demonstrate this, we consider a domain wall between the two MPCs with opposite inversion-symmetric unit cells (see Fig.~1(e) in the main text).
Thus, the defect divides the crystal into two regions described by unit cells with opposite inversion centers. 

\begin{figure}[h]
    \centering
    \includegraphics[width=0.40\linewidth]{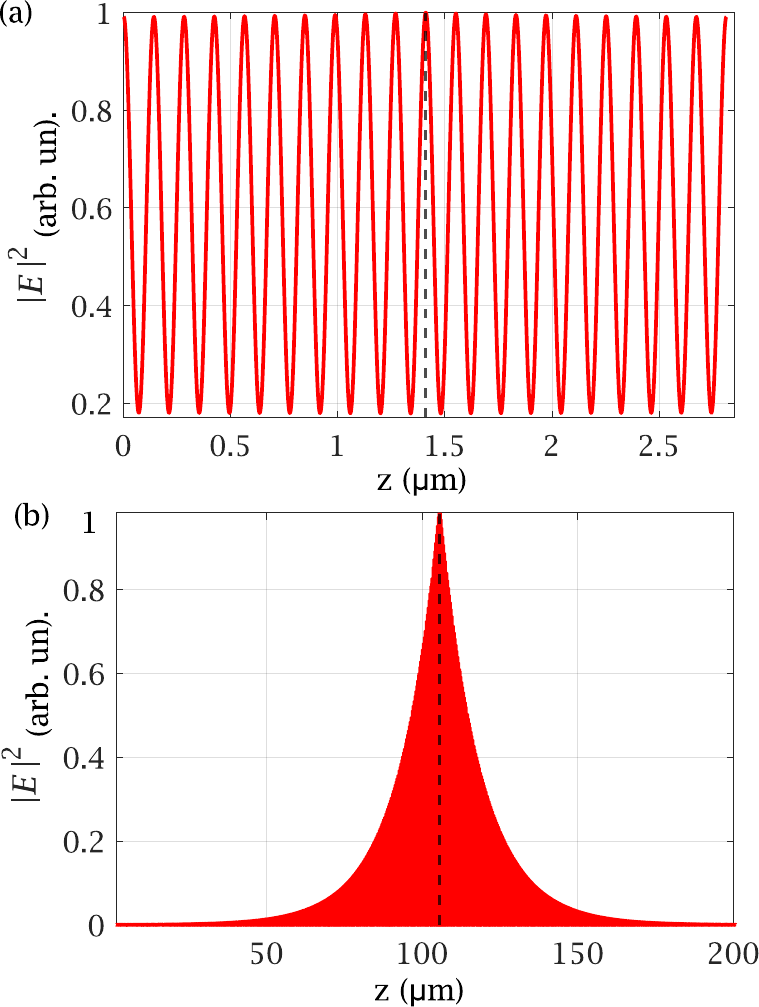}
    \caption{Spatial distribution of topological defect mode arising at the boundary of two MPCs. $z$ axis is normal to the layers.  Interface mode profile $|E|^2$ for the finite structure of (a) $N=20$ periods. Localization is hardly visible due to the small size of the sample; (b) $N=1500$ periods. The localization length of the topological defect mode $1/\kappa\approx26$~$\mu$m. Due to symmetry, the frequencies and spatial profiles of LCP and RCP topological modes coincide. Parameters: $g=0.03$, $\eps_1=\eps_2=5.5$, $d=70$~nm.   
    }
    \label{fig:modes}
\end{figure}

A characteristic parameter of this structure is the decay length of the localized state $1/\kappa$. 
Inside the gap the field is evanescent, $k=\pi/a+i\kappa$, so $|E(z)|\propto e^{-\kappa z}$ and the field amplitude decays exponentially. 
To find the frequency of the topological states for samples with large enough length $L \gtrsim \kappa^{-1}$, we search for the frequency for which the impedances for the corresponding semi-infinite samples are opposite in sign, in accordance with the theory in Ref.~\cite{Xiao2014Apr}. 
In terms of surface impedances in circular basis and taking into account the opposite normals of the two concatenated surfaces, the surface state appears for the frequency $\omega_s$ such that $Z^- (\omega_s) - Z^+ (\omega_s) = 0$, i.e. the impedances are matched for a mode with a certain circular polarization. 
Note that such structures with inversion-symmetric unit cells feature a termination different from the conventional $AB$-stacking, and thus surface impedances are generally different from those found earlier in Eqs.~\eqref{impedances_midgap_AB}. At the bandgap center, re-calculating impedances to leading order or these structures gives 
\begin{equation} \label{impedances_midgap_AB}
    Z^- (\omega_B) = - (1+ \mu) / \bar n + O(\mu^2) , \qquad 
    Z^+ (\omega_B) = - (1- \mu) / \bar n + O(\mu^2) ,
\end{equation}
with $ Z^- (\omega_B) - Z^+ (\omega_B) = 2 \mu / \bar n + O(\mu^3) $. Thus, to zeroth order in $\mu$, $\omega_s \simeq \omega_B$, so the localized states appear close to, but not exactly at the frequency of the bandgap center. 

The corresponding localization length of the topological states is then well-approximated by the infinite-crystal imaginary part of the wavenumber at the midgap, Eq.~\eqref{kappa__a}, 
\begin{equation} \label{loc_length}
    \ell \equiv \kappa^{-1} \simeq a / (2\mu) \simeq \varepsilon a/ g 
    . 
\end{equation}

\begin{figure}[!b]
    \centering
    \includegraphics[width=0.8\linewidth]{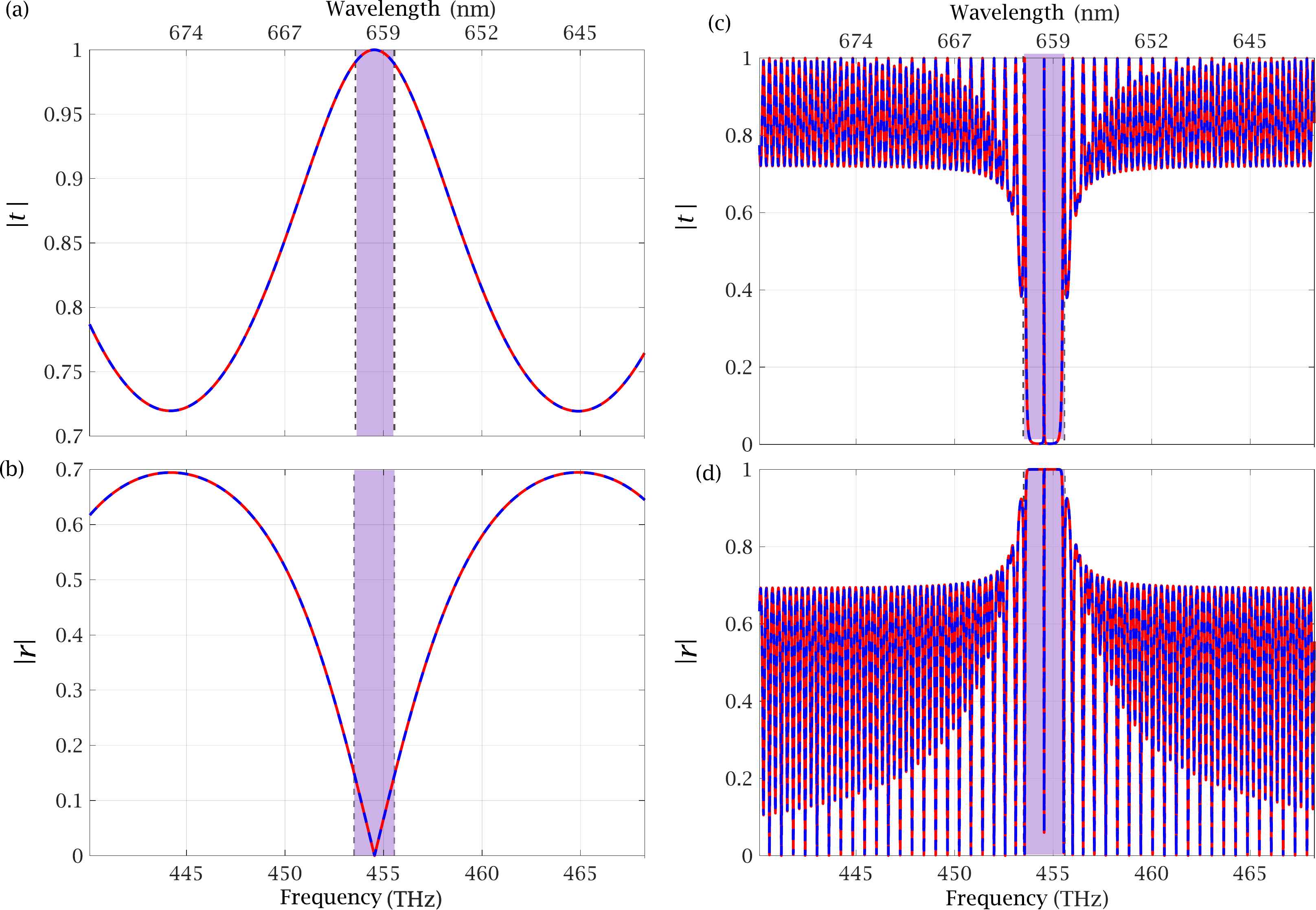}
    \caption{Resonant transmission (a) and reflection (b) for RCP (blue line) and LCP modes (red line), N=20. For such number of periods the localization is not seen enough. For N=1500 one may see an explicit peak of the transmittance (c) and dip of the reflectance (d).}
    \label{fig:topzones}
\end{figure}

To find the exact profiles of the localized modes, we solve the more bulky eigenvalue problem without incident field, since the localized mode is the eigenmode of the finite crystal. Taking into account that the localized mode corresponds to the solution without sources, i.e. $E^{\eta}_i=0$, this can be written in the matrix form (using the full finite-structure transfer-matrix) as 
\begin{equation}
\begin{pmatrix}
E^t_{\eta}\\0
 \end{pmatrix}
    =\hat{M}_{total,\eta}
    \begin{pmatrix}
        0 \\E^r_{\eta}
    \end{pmatrix}
    . 
\end{equation}
The matrix equation then yields:
\begin{align}
    E^{\eta}_t=M^{\eta}_{12}E^{\eta}_r,\qquad 0=M^{\eta}_{22}E^{\eta}_r. 
\end{align}
A nontrivial solution requires $M_{22} (\omega_s)=0$, which determines the exact frequency $\omega_s$ of the localized mode. 
At the same time, the field profiles are obtained by propagating the vector $v_0 = (0,1)^{\rm T}$ by $v_j = P_j(\omega_s) v_{j-1}$, where $P_j(\omega_s)$ is the $j^{\rm th}$ transfer-matrix block.

For example, for a photonic crystal with $\eps=5.5$, $g=0.03$, $d=70 $~nm and number of the periods $N=20$ the decay length is $\kappa^{-1} \simeq 25.7 \,\mu m$ 
is much larger than the total size  of the structure $L=2Nd=2.8\mu m$. Consequently, the field cannot decay sufficiently within the structure, and the mode does not become well localized, see Fig.~\ref{fig:modes}(a). For a finite crystal, it acts as a resonator for the defect state: multiple reflections within the crystal cause interference of the defect field amplitude. 
The edge state is well-localized for larger unit cells, see the example in Fig.~\ref{fig:modes}(b).

The localized topological state manifests itself also in the reflection and transmission properties of MPC: it enables the reflection dip and transmission peak near the center of the bandgap, see Fig.~\ref{fig:topzones}. 


\section{Ellipticity and polarization rotation}

In this section, we derive the polarization rotation angle and ellipticity to describe the polarization state of the scattered light. We consider a general elliptically polarized wave. In Cartesian coordinates, the projections of the complex electric field amplitude read:
\begin{align}
    &E_x=E_{0x}\,e^{i\delta-i\omega t},\qquad E_y=E_{0y}\,e^{-i\omega t}\:,
\end{align}
where without loss of generality, we set the initial phase of $E_y$ to zero and $\delta$ measures the phase delay between the $x$ and $y$ components of the field. Taking the real part of the above complex amplitudes, we recover the real time-dependent fields
\begin{equation}
E_x(t)=E_{0x}\,\cos(\om t-\delta),\qquad E_y(t)=E_{0y}\,\cos \om t\:.
\end{equation}
Excluding time $t$ from this parametric equation, we recover 
\begin{equation}
\frac{E_x^2(t)}{E^2_{0x}} +\frac{E_y^2(t)}{E^2_{0y}}-\frac{2E_x(t)\,E_y(t)}{E_{0x}E_{0y}}\cos\delta=\sin^2\delta \label{defaultellipse}.
\end{equation}
The above equation describes an ellipse in $E_x-E_y$ plane, while the angle $\delta$ is expressed in terms of complex amplitudes as follows:
\begin{gather}
 \cos\delta=\frac{\text{Re}(E_xE^*_y)}{E_{0x}\,E_{0y}}\:,\\
 \sin\delta=\frac{\text{Im}(E_xE^*_y)}{E_{0x}\,E_{0y}}\:.
\end{gather}
Equation~\eqref{defaultellipse} can be brought to the canonical form 
\begin{equation}
\frac{E^2_{x'}}{E^2_{0x'}}+\frac{E^2_{y'}}{E^2_{0y'}}=1
\label{strightellipse}
\end{equation}
by rotating the coordinate frame by some angle $\theta$ via the rotation matrix
\begin{equation}
\hat{R}=
\begin{pmatrix}
\cos(\theta)&&\sin(\theta)\\
-\sin(\theta)&& \cos(\theta)
    \end{pmatrix},
    \end{equation}
so that
\begin{equation}
\begin{pmatrix}
E_{x'}\\
E_{y'}
\end{pmatrix}
=\hat{R}\,
\begin{pmatrix}
E_{x}\\
E_{y}
\end{pmatrix}\:.
\end{equation}
We now choose the rotation angle $\theta$ such that the general equation~\eqref{defaultellipse} is converted to the canonical form~\eqref{strightellipse}, which yields
\begin{equation}
\tan (2\,\theta)=\frac{2\,E_{0x} E_{0y}\,\cos\delta}{E_{0x}^2-E_{0y}^2}=\frac{2\,\text{Re}(E_{x}E^*_y)}{E_{0x}^2-E_{0y}^2}
\label{rotationpolarization}.
\end{equation}
Note that this formula does not distinguish whether polarization rotation is equal to $\theta$ or $\pi/2+\theta$, and the choice should be made based on the other considerations, e.g. continuity. Furthermore, if $E_{0x}=E_{0y}$, Eq.~\eqref{rotationpolarization} features the singularity. If the phase shift between $E_x$ and $E_y$ complex amplitudes is different from $\pm\pi/2$, this yields $\tan\,2\theta\rightarrow\infty$, i.e. $\theta=\pi/4$. Otherwise, if $E_x$ and $E_y$ are shifted by $\pm\pi/2$, the polarization of the wave is circular and hence polarization plane is not defined.

Yet another characteristic of the wave is ellipticity $\eta$ which is related to the ratio of the two components of polarization ellipse. In the general case it is introduced by the expression
\begin{equation}
\sin(2\eta)=\frac{2\,\text{Im}(E_{x}E^*_y)}{|E_{0x}|^2+|E_{0y}|^2}.
    \label{ellipticity}
\end{equation}
For instance, $\eta=0$  corresponds to the linear polarization, while  $\eta=\mp\frac{\pi}{4}$ captures right and left circular polarizations, respectively. All other nonzero $\eta$ from the interval $(-\pi/4,\pi/4)$ capture general elliptical polarization of the wave.

\section{Antiferromagnetic MPC in a dielectric environment}


\begin{figure}[h]
    \centering
    \includegraphics[width=0.85\linewidth]{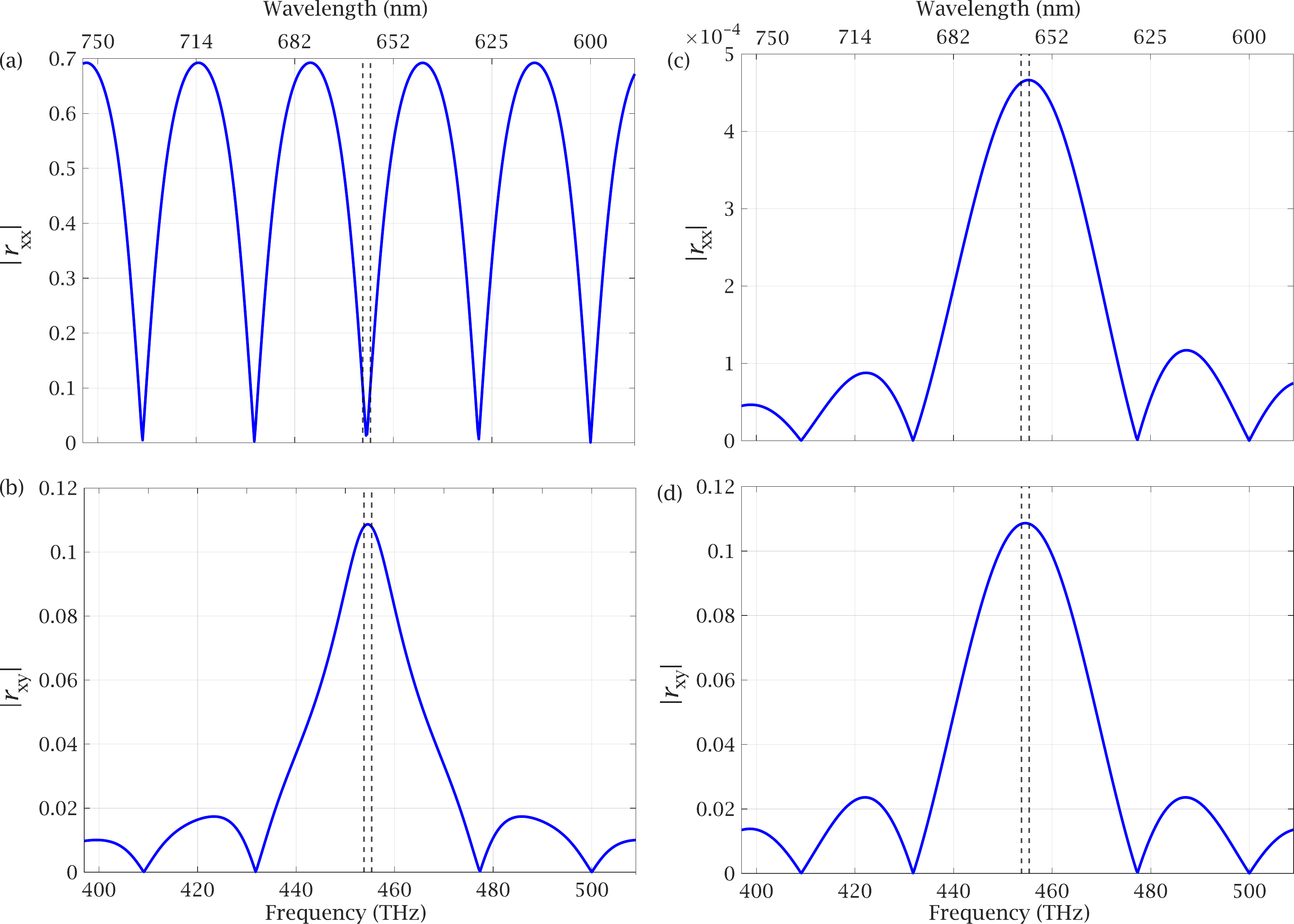}
        \caption{
        Co-polarized~(a) and cross-polarized~(b) reflection coefficients for the finite slab with $N=20$, $g=0.03$, $\eps=5.5$, $d=70$~nm surrounded by the vacuum with $\varepsilon_a = 1$ and co-polarized (c) and cross-polarized (d) reflection coefficients for the finite slab surrounded by the infinite ambient medium with $\varepsilon_a = 5.5$.  Cross-polarized reflection is practically insensitive to the ambient refractive index.
        }
    \label{fig:rxx bulk}
\end{figure}

Here we consider the finite AFM MPC surrounded by the infinite ambient medium with $\varepsilon_a = 5.5$ matched to that of the layers of the MPC, see Fig.~\ref{fig:rxx bulk}. 
We observe that, while co-polarized $r_{xx}$ dramatically diminishes when the ambient medium has the same $\varepsilon$ as the layers themselves, 
the cross-polarized reflection amplitude $|r_{xy}| \simeq 0.1$ remains practically unchanged 
{in comparison with air-structure-air [no substrate] case, see the curves for the same $g=0.03$ and other parameters Fig.~3 in the main text.} 
This could be explained by analytically examining the Bloch impedance at the $\omega_B$ frequency of the first bandgap center. 

To that end, we switch to the basis of electric and magnetic field amplitudes \((E_\eta,H_\eta)\), \(\eta=\pm1\). A forward wave in a nonmagnetic medium of index \(n_0\) has
\[
H_\eta^+=-i\eta n_0 E_\eta^+,
\qquad
Z_{0,\eta}\equiv \frac{E_\eta^+}{H_\eta^+}
=
\frac{i\eta}{n_0}.
\]
Further, if the finite stack seen from the left has input impedance \(Z_{{\rm in},\eta}\), the reflection coefficient for the circular polarization reads
\[
r_\eta=
\frac{Z_{{\rm in},\eta}-Z_{0,\eta}}
     {Z_{{\rm in},\eta}+Z_{0,\eta}} .
\]
Consider the layer stack \((AB)^N\) comprising $N$ unit cells at the frequency of the center of the first bandgap. 
As established in Section S2, the one-cell transfer matrix for a particular circular polarization $\eta$ has eigenvalues
$
\lambda=-e^{-\kappa_\eta a}, \, 
\lambda^{-1}=-e^{+\kappa_\eta a}
$
at the bandgap center equal for both RCP and LCP if the layers have no additional permittivity contrast. The corresponding (unnormalized) Bloch eigenvectors in the same basis \((E_\eta,H_\eta)\) are proportional to 
\[
v_{d, \eta}=\binom{Z_{d, \eta}}{1},
\qquad
v_{g, \eta}=\binom{Z_{g, \eta}}{1},
\]
where \(d\) and \(g\) denote decaying and growing under propagation to the right, and $Z_{d, \eta}$ and $Z_{g, \eta}$ are corresponding Bloch impedances. Then, if at the left edge of the stack, 
\[
\binom {E_\eta} {H_\eta}_{z=0}=A v_{d,\eta}+B v_{g,\eta},
\]
then at the right edge
\[
\binom {E_\eta} {H_\eta}_{z=Na}
=
A\lambda^N v_{d,\eta}+B\lambda^{-N}v_{g,\eta} .
\]
If the right exterior has impedance \(Z_{0,\eta}\), the transmitted wave satisfies
\[
\frac {E_\eta} {H_\eta} =Z_{0,\eta}.
\]
Therefore
\[
A\lambda^N(Z_{d,\eta}-Z_{0,\eta})
+
B\lambda^{-N}(Z_{g,\eta}-Z_{0,\eta})=0,
\]
so
\[
\frac BA
=
-\lambda^{2N}
\frac{Z_{d,\eta}-Z_{0,\eta}}{Z_{g,\eta}-Z_{0,\eta}}.
\]
Hence the input impedance of the finite AFM stack is
\begin{equation} \label{Z_AFM_eta}
Z_{{\rm AFM},\eta}^{(N)}
=
\frac{Z_{d,\eta}+\beta_\eta Z_{g,\eta}}{1+\beta_\eta},
\qquad
\beta_\eta=
-\lambda^{2N}
\frac{Z_{d,\eta}-Z_{0,\eta}}{Z_{g,\eta}-Z_{0,\eta}}.
\end{equation}

As established earlier, for weak gyrotropy, $\kappa a\simeq 2\mu$ ($\mu\simeq \frac{g}{2\varepsilon}$), while the Bloch impedances are 
$
Z_{d,+}\sim \mu^{-1}, Z_{g,+}\sim \mu,
$
and
$
Z_{d,-}\sim \mu, Z_{g,-}\sim \mu^{-1}.
$
Thus, provided
\[
Z_{d,+}\gg |Z_{0,+}|\gg |Z_{g,+}|,
\qquad
|Z_{g,-}|\gg |Z_{0,-}|\gg |Z_{d,-}|,
\]
one obtains, with
\[
s=N\kappa a,
\qquad
\rho=\tanh s,
\]
the leading finite-stack impedances
\[
Z_{{\rm AFM},+}^{(N)}
\simeq
Z_{0,+}e^{2s}
=
Z_{0,+}\frac{1+\rho}{1-\rho},
\]
and
\[
Z_{{\rm AFM},-}^{(N)}
\simeq
Z_{0,-}e^{-2s}
=
Z_{0,-}\frac{1-\rho}{1+\rho}.
\]
Substitution into the reflection formula gives
\begin{equation}
r_+
\simeq
\frac{e^{2s}-1}{e^{2s}+1}
=
\tanh s
=
\rho,
\end{equation}
whereas
\begin{equation}
r_-
\simeq
\frac{e^{-2s}-1}{e^{-2s}+1}
=
-\tanh s
=
-\rho.
\end{equation}
Therefore, at the gap center,
\begin{equation}
r_{xx}
=
\frac{r_++r_-}{2}
\simeq0,
\end{equation}
while
\begin{equation}
r_{xy}
=
\frac{i(r_- - r_+)}{2}
\simeq
-i\rho
=
-i\tanh(N\kappa a).
\end{equation}
Thus we obtain a remarkably simple formula 
\begin{align}
\label{pur2}
|r_{xy}|
\simeq
\tanh(N\kappa a) \simeq \tanh(L / \ell) \simeq 
\tanh \left[ (L/a) \cdot (g / \varepsilon) \right] = \tanh \left[ N g / \varepsilon \right]
,
\end{align}
with $\ell \equiv \kappa^{-1}$ is obtained from Eq.~\eqref{loc_length}, and the leading result $\theta \simeq \pi/2 + O(g)$ is independent of the exterior impedance \(Z_{0,\eta}\). 
Thus, the midgap cross-polarized Bragg reflection distribution into co- and cross-polarized channels is indeed essentially unchanged when the exterior is changed from air to a dielectric environment.

To obtain the leading-order correction to the rotation angle $\theta \simeq \pi/2$, we need to keep next-order terms in respective impedances. 
As we will see, this correction is somewhat sensitive to ambient refractive index, although it is still linear in small $g \ll 1$. 

From earlier equations~\eqref{Z_+_CB}-\eqref{Z_-_CB}, we obtain the Bloch impedances to leading order, 
\begin{equation}
Z_{d,+}=Z_{g,-}
=
\frac{P}{\mu}+O(\mu),
\qquad
Z_{g,+}=Z_{d,-}
=
Q\mu+O(\mu^3),
\end{equation}
where we introduced the notations 
\begin{equation}
P=\frac{4}{\pi\bar n},
\qquad
Q=-\frac{\pi}{4\bar n},
\qquad
PQ=-\frac{1}{\bar n^2}.
\end{equation}
We also introduce for convenience 
$
t=e^{-2s} = \lambda^{2 N} = e^{- 2 N g / \varepsilon},
$
and write the ambient impedance in the circular basis as
\[
Z_{0,\eta}=i\eta z_a,
\qquad
z_a=\frac{1}{n_a}.
\]
Substituting the Bloch impedances into Eqns.~\eqref{Z_AFM_eta} and expanding to first order in $\mu$, we obtain 
\begin{align}
r_+
&=
\rho+
\frac{2i\mu(t-1)\left(z_a^2-PQt\right)}
     {Pz_a(1+t)^2}
+O(\mu^2),\\
r_-
&=
-\rho+
\frac{2i\mu(t-1)\left(tz_a^2-PQ\right)}
     {Pz_a(1+t)^2}
+O(\mu^2), 
\end{align}
which now contain leading-order corrections. Thus,
\begin{align}
r_{xx}
=
-i\mu\rho\,
\frac{z_a^2-PQ}{Pz_a}
+
O(\mu^2)
=
-i\frac{\pi\mu}{4}
\left(
\frac{n}{n_a}+\frac{n_a}{n}
\right)
\tanh(N\kappa a)
+
O(\mu^2)
=
-i\frac{\pi}{8}
\left(
\frac{n}{n_a}+\frac{n_a}{n}
\right)
\frac{g}{\varepsilon}
\tanh\!\left(N\frac{g}{\varepsilon}\right)
+
O(g^2)
.
\end{align}
%
The polarization rotation angle of the reflected light is thus 
\begin{equation}
\theta
=
\frac{\pi}{2}
+
\frac{\pi}{8}
\left(
\frac{n}{n_a}+\frac{n_a}{n}
\right)
\frac{g}{\varepsilon}
+O(g^2)
.
\end{equation}

\section{Comparison of antiferromagnetic and ferromagnetic MPC}

In this section, we compare the strength of magneto-optical effects for antiferromagnetic (AFM) and ferromagnetic (FM) arrangements of the same gyrotropic layers placed on the same substrate.

\begin{figure}[h]
    \centering
    \includegraphics[width=0.4\linewidth]{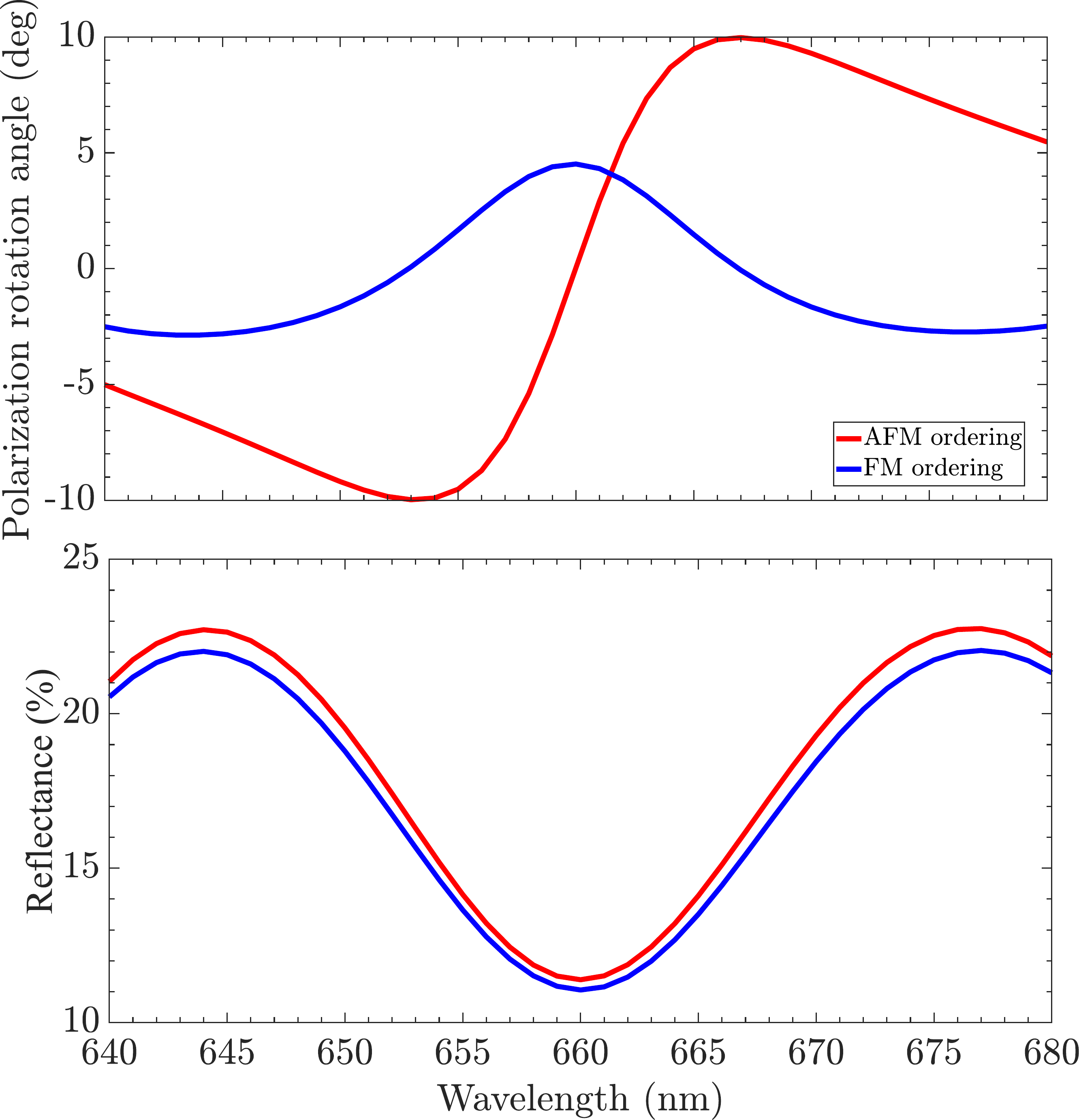}
    \caption{Polarization rotation and reflectance  in the cases of ferromagnetic (FM, blue line) and antiferromagnetic (AFM, red line) structures placed on gadolinium gallium garnet substrate. In FM case we consider a single uniformly magnetized  magnetic film of thickness $2Nd$. The simulation is performed for parameters $N=20$ layers, $g=0.03$, $\eps=5.5$, $d=70$~nm. The geometry of the system is  air/structure/substrate with the substrate permittivity $\varepsilon=4.0$. The rotation angle in AFM configuration is nearly twice larger than that in FM case.}
    \label{fig:comparison_AFM_FM}
\end{figure}

\begin{figure}[h]
    \centering
    \includegraphics[width=0.4\linewidth]{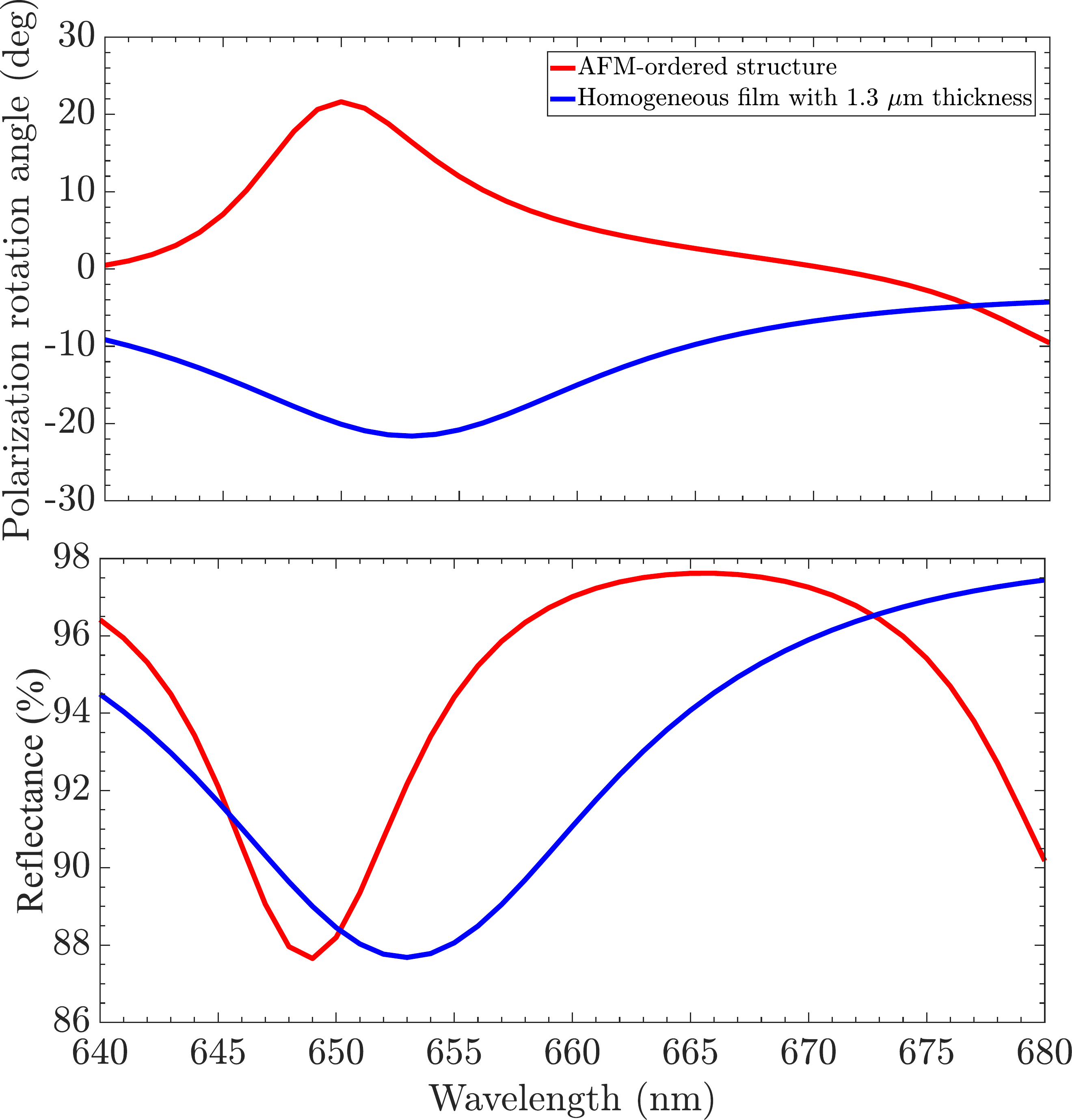}
    \caption{Polarization rotation and reflectance  in the cases of ferromagnetic (FM, blue line) and antiferromagnetic (AFM, red line) structures placed on the silver substrate. The permittivity of gyrotropic layers $\eps=5.5$, gyrotropy $g=0.03$, number of periods in AFM structure is $N=20$, so that the total thickness is 2.8~$\mu$m. The thickness of the respective ferromagnetic film is only 1.3~$\mu$m.}
    \label{fig:comparison_AFM_FM_new}
\end{figure}

First we examine the case of gadolinium gallium garnet (GGG) substrate modeled as a dielect-ric with permittivity $\eps=4$ [Fig.~\ref{fig:comparison_AFM_FM}. We observe that the reflection from antiferromagnetic structure is slightly higher than that from the  ferromagnetic layer which should be attributed to the bandgap in the dispersion of antiferromagnetic film. However, the bandgap is narrow, and hence the increase is quite marginal, so that the reflection stays close to 20\%. The gain in polarization rotation is more pronounced, being twice stronger for AFM configuration (nearly $10^\circ$) compared to the FM one (nearly $5^\circ$ at maximum). Thus, in the case of a dielectric substrate the designed antiferromagnetic structure improves reflectance because of the photonic bandgap and provides larger values of polarization rotation.

Next we examine an alternative scenario when the structure is placed on a silver mirror [Fig.~\ref{fig:comparison_AFM_FM_new}]. Conducting silver substrate boosts the the reflection up to 90\%. In such setting, however, AFM configuration features no advantages compared to the single FM film: due to the conducting substrate, the presence of photonic bandgap does not play a defining role. We observe that the comparable $20^\circ$ rotation in reflection is achieved for 2.8~$\mu$m AFM structure with $N=20$ lattice periods and only 1.3~$\mu$m-thick FM structure, i.e. FM structure appears nearly twice more efficient.


\section{The role of material losses}

In this section, we analyze the effect of losses on polarization rotation and reflectance of the designed AFM magnetophotonic crystal. We examine the case  bismuth iron garnet thin films, whose optical properties are taken from the experimental data of Ref.~\cite{wittekoek1975magneto} at a wavelength of 645 nm. This wavelength was intentionally chosen as it lies within a transparency window of the material, ensuring that both the permittivity and gyration tensor components exhibit only minor imaginary parts. The adopted dielectric permittivity and gyration values are
$\varepsilon=5.5+0.0849\,i$ and 
$g=0.03+0.0026\,i$, respectively.

\begin{figure}[h]
    \centering
    \includegraphics[width=0.4\linewidth]{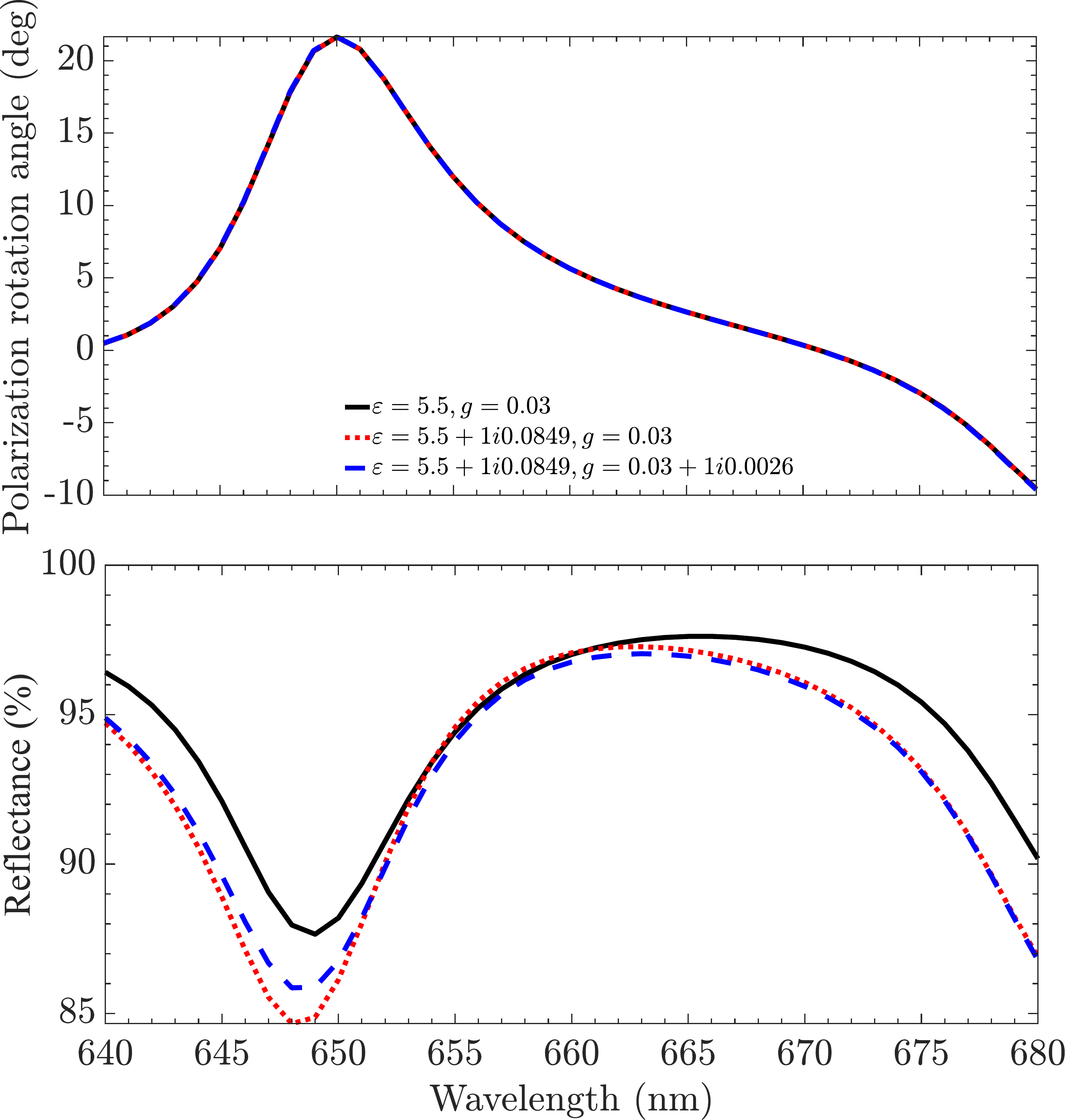}
    \caption{Wavelength dependence of the polarization rotation angle  and reflectance  for the three distinct material configurations of the proposed AFM multilayer structure (air / AFM stack /
silver): (1) both the permittivity tensor and the gyration are taken as purely real for each layer; (2) mixed case – the permittivity tensor is complex (accounting for absorption losses), while the gyration remains purely real; and (3) full-complex case – both the permittivity and gyration tensors are complex, incorporating simultaneous dissipative and non-Hermitian gyrotropic effects. The comparison elucidates the separate and combined roles of dielectric losses and complex gyrotropic response in shaping the spectral features of magneto-optical effects.}
    \label{fig: imag real complex param}
\end{figure}

In Fig.~\ref{fig: imag real complex param}, we compare the results for the system with realistic losses and for the idealized lossless system. The analysis reveals that the inclusion of the imaginary parts of both 
$\varepsilon$ and 
$g$ has a negligible effect on the calculated Faraday rotation angle. However, these losses do lead to a moderate reduction in the overall reflectance, decreasing it by approximately 5\% relative to the lossless case. This minor quantitative impact confirms that the results obtained for lossless AFM MPCs are relevant for the experimental situation.

\bibliography{PBG}